\documentstyle[12pt,epsf]{article}
\newcommand{\be}{\begin{equation}}
\newcommand{\ee}{\end{equation}}

\newcommand{\beqa}{\begin{eqnarray}}
\newcommand{\eeqa}{\end{eqnarray}}
\newcommand{\nn}{\nonumber}

\newcommand{\eqref}[1]{(\ref{#1})}

\def\boxit#1{\vbox{\hrule\hbox{\vrule\kern8pt
\vbox{\hbox{\kern8pt}\hbox{\vbox{#1}}\hbox{\kern8pt}}
\kern8pt\vrule}\hrule}}
\def\mathboxit#1{\vbox{\hrule\hbox{\vrule\kern8pt\vbox{\kern8pt
\hbox{$\displaystyle #1$}\kern8pt}\kern8pt\vrule}\hrule}}

\def\IB{\relax\hbox{$\inbar\kern-.3em{\rm B}$}}
\def\IC{\relax\hbox{$\inbar\kern-.3em{\rm C}$}}
\def\ID{\relax\hbox{$\inbar\kern-.3em{\rm D}$}}
\def\IE{\relax\hbox{$\inbar\kern-.3em{\rm E}$}}
\def\IF{\relax\hbox{$\inbar\kern-.3em{\rm F}$}}
\def\IG{\relax\hbox{$\inbar\kern-.3em{\rm G}$}}
\def\IGa{\relax\hbox{${\rm I}\kern-.18em\Gamma$}}
\def\IH{\relax{\rm I\kern-.18em H}}
\def\IK{\relax{\rm I\kern-.18em K}}
\def\IL{\relax{\rm I\kern-.18em L}}
\def\IP{\relax{\rm I\kern-.18em P}}
\def\IR{\relax{\rm I\kern-.18em R}}
\def\IZ{\relax\ifmmode\mathchoice
{\hbox{\cmss Z\kern-.4em Z}}{\hbox{\cmss Z\kern-.4em Z}}
{\lower.9pt\hbox{\cmsss Z\kern-.4em Z}} {\lower1.2pt\hbox{\cmsss
Z\kern-.4em Z}}\else{\cmss Z\kern-.4em Z}\fi}

\def\II{\relax{\rm I\kern-.18em I}}

\def\CD {{\cal D}}

\def\CH {{\cal H}}

\def\CL {{\cal L}}

\begin{document}

\begin{flushright}
NRCPS-HE-09-55\\
August, 2009
\end{flushright}

\vspace{1cm}
\begin{center}

{\Large \it Particle Spectrum  \\
\vspace{0,3cm} of  \\
\vspace{0,3cm}
Non-Abelian Tensor Gauge Fields
} 

\vspace{1cm}

{ \it{George  Savvidy  } }

\vspace{0.5cm}

 {\it Institute of Nuclear Physics,} \\
{\it Demokritos National Research Center }\\
{\it Agia Paraskevi, GR-15310 Athens, Greece} 
\end{center}
\vspace{60pt}

\centerline{{\bf Abstract}}

\vspace{12pt}

\noindent
We present a brief review of the non-Abelian tensor gauge field theory and analyze its
free field equations for lower rank gauge fields when the interaction coupling constant
tends to zero. The free field equations are written in terms of the first order derivatives
of extended field strength tensors similar to the electrodynamics and non-Abelian gauge
theories. We determine the particle content of the free field equations and count the
propagating modes which they describe. In four-dimensional space-time the rank-2 gauge
field describes propagating modes of helicity two and zero. We show that the rank-3
gauge field describes propagating modes of helicity-three and a doublet of helicity-one
gauge bosons. Only four-dimensional space-time is physically acceptable, because in
five and higher-dimensional space-time the equation has solutions with negative norm
states. We discuss the structure of the particle spectrum for higher rank gauge fields.

\vspace{150 pt}

\centerline{\it Dedicated to Professor Emmanuel Floratos at the occasion of his 60th birthday}


\newpage

\pagestyle{plain}

\tableofcontents

\section{\it Introduction}

It is well understood that the concept of local gauge
invariance allows to define non-Abelian gauge field \cite{yang,chern},
to derive its dynamical field equations
and to develop a universal point of view on matter interactions
as resulting from the exchange of gauge quanta of spin one.
It is appealing to extend the gauge principle so that it would define the
interaction of matter fields which carry not only non-commutative internal charges,
but also arbitrary spins \cite{Savvidy:2005fi,Savvidy:2005zm,Savvidy:2005ki}.

In our recent approach  the gauge fields are defined as
rank-$(s+1)$ tensors
\cite{Savvidy:2005fi,Savvidy:2005zm,Savvidy:2005ki}
$$
A^{a}_{\mu\lambda_1 ... \lambda_{s}},
$$
and they are totally symmetric with respect to the
indices $  \lambda_1 ... \lambda_{s}  $. The number of symmetric
indices $s$ runs from zero to infinity
\footnote{  {\it A priori} the tensor fields
have no symmetries with respect to the first index  $\mu$.
The  $a$ is the adjoint index. The totally symmetric tensors were
considered in
\cite{fierz,fierzpauli,yukawa1,wigner,schwinger,
Weinberg:1964cn,chang,singh,fronsdal,berends,Bengtsson:1983pd,Guttenberg:2008qe}.}.
The first member of this family of tensor gauge fields is the Yang-Mills
vector boson $A^{a}_{\mu}$.
The extended non-Abelian gauge transformation $
\delta_{\xi} $ of tensor gauge fields
is defined
by the equation (\ref{polygauge}) and
comprises a closed algebraic structure \cite{Savvidy:2005fi,Savvidy:2005zm,Savvidy:2005ki}.
This allows to define generalized field strength tensors
\cite{Savvidy:2005fi,Savvidy:2005zm,Savvidy:2005ki}
$$
G^{a}_{\mu\nu,\lambda_{1}...\lambda_{s}}
$$
which are transforming   homogeneously
with respect to the extended gauge transformations $\delta_{\xi} $.
Using these field strength tensors one can
construct two infinite series of quadratic forms
$
{{\cal L}}_{s}$ and ${{\cal L}}^{'}_{s}~(s=2,3,...)
$
invariant with respect to the
transformations $\delta_{\xi} $
\cite{Savvidy:2005fi,Savvidy:2005zm,Savvidy:2005ki}. These forms
contain quadratic kinetic terms, as well as cubic and  quartic terms  describing
nonlinear interaction of gauge fields with dimensionless
coupling constant $g$.
In order to make all tensor gauge fields dynamical one should add all these
forms in the Lagrangian.

Thus the gauge invariant
Lagrangian describing dynamical tensor gauge fields of all ranks
has the form \cite{Savvidy:2005fi,Savvidy:2005zm,Savvidy:2005ki,Savvidy:2005vm,Savvidy:2006qf}
\be\label{generalgaugedensity}
{{\cal L}} = {{\cal L}}_{YM} + g_{2}{{\cal L}}_{2}+ g^{'}_{2}{{\cal L}}^{'}_{2}
+g_{3}{{\cal L}}_{3}+ g^{'}_{3}{{\cal L}}^{'}_{3} +...
\ee
The coupling constants $g_{s}$ and $g^{'}_{s}~(s=2,3,...)$ remain arbitrary
because each term is separately invariant
with respect to the extended gauge transformations $\delta_\xi$
leaves these coupling constants undetermined.
Our aim is to analyze the particle spectrum of the theory and to
define coupling constants $g_{s}$ and $g^{'}_{s}$ at which there are no negative norm
states in the spectrum.
Tensor gauge fields have many components, some of them define physical
propagating modes of definite helicities, and some may correspond to unphysical negative
norm states.
As we shall see, only for specific values of the coupling constants
$g^{'}_{s}$ and only in four-dimensional space-time
the particle spectrum becomes completely physical.

Let us consider first a linear sum of two gauge invariant forms in (\ref{generalgaugedensity})
$$
g_{2}{{\cal L}}_{2}+ g^{'}_{2}{{\cal L}}^{'}_{2}~,
$$
which defines the kinetic operator and nonlinear interactions of the rank-2 tensor gauge
field $A^{a}_{\mu\lambda}$\footnote{It has sixteen components in the four-dimensional space-time.}.
As we found in
\cite{Savvidy:2005fi,Savvidy:2005zm,Savvidy:2005ki,Konitopoulos:2007hw},
if one chooses the coupling constant
$g^{'}_{2}=g_2$, then the sum
exhibits invariance with respect to
a bigger gauge group. In that case the free field equation (\ref{freeequations1})/(\ref{mainequation})
for the rank-2 tensor gauge field
describes the propagation of {\it helicity-two, $\lambda = \pm 2 $, and
helicity-zero, $\lambda =  0$, massless charged
tensor gauge bosons} and there are no propagating negative norm states.
This result will be recapitulated in the third section.

Our aim here is to extend this analysis to the rank-3 tensor gauge
field. Considering the linear sum
$$
g_{3}{{\cal L}}_{3}+ g^{'}_{3}{{\cal L}}^{'}_{3}
$$
we shall demonstrate that if one chooses the coupling
constant $g^{'}_{3} = {4 \over 3} g_3$ then the sum
again exhibits invariance with respect to
a bigger gauge group \cite{Savvidy:2006qf}.  The explicit description of this symmetry
together with the corresponding free field equation (\ref{freethirdrankequations})/
(\ref{freeequation})
for the rank-3 tensor gauge field $A_{\mu\lambda_1\lambda_2}$ will be given
in the forth and fifth sections\footnote{Its relation to the Schwinger
equation for the symmetric rank-3 tensor gauge field is discussed
in the fifth section. See also references
\cite{schwinger,Savvidy:2006qf,Guttenberg:2008qe}. }.
We shall demonstrate that in {\it four-dimensional space-time} the free
equation (\ref{freethirdrankequations})/
(\ref{freeequation})
describes the {\it propagation of helicity-three, $\lambda = \pm 3 $, and
a doublet of helicity-one,
$\lambda = \pm 1,\pm 1$, massless charged gauge bosons} and that
there are no propagating negative norm states.  {\it The four-dimensional
space-time is  critical because in five- and higher-dimensional space-time the
equation has solutions with negative norm states.}

Summarizing our findings we can state that the Lagrangian $\CL$ describes the interacting
system of gauge bosons of increasing helicities. The system has
Yang-Mills gauge boson on the first level (s=0), the helicity-two and -zero gauge bosons
on the second level (s=1) and the helicity-three and a doublet of
helicity-one gauge bosons on the third level (s=2).

The particle spectrum on higher levels is not yet known completely and
to find it out  remains a challenging
problem. The problem consists in finding out the
value of the coupling constant  $g^{'}_{s+1}$ at which
the corresponding free field equation for the rank-(s+1) gauge field is free from
propagating negative norm states. As we have found for
$$
g^{'}_{s+1} = {2s \over s+1} g_{s+1},~~~~~~s=0,1,2,...
$$
($g_1=g_{YM})$ there are two solutions which describe the propagating
positive norm states of helicities $\lambda =\pm (s+1)$.
But the difficulty in finding out {\it all} propagating modes for this value  of the
coupling constant $g^{'}_{s+1}$ lies in the fact that the number of
field components dramatically
increases with the rank of the tensor gauge field: in the case of rank-2 gauge field
we had sixteen components and in the case considered in this article for
the rank-3 gauge field we have to analyze an equation with forty components.
The presented analysis shows that, most probably, the full system is
unitary for all higher-rank non-Abelian tensor gauge fields and
only in four-dimensional space-time.

First let us recapitulate the general form of the Lagrangian ${{\cal L}}$  in (\ref{generalgaugedensity}).

\section{\it Non-Abelian Tensor Fields}

The gauge fields are defined as rank-$(s+1)$ tensors
\cite{Savvidy:2005fi,Savvidy:2005zm,Savvidy:2005ki}
$$
A^{a}_{\mu\lambda_1 ... \lambda_{s}}(x),~~~~~s=0,1,2,...
$$
and are totally symmetric with respect to the
indices $  \lambda_1 ... \lambda_{s}  $.  The index $a$ numerates the generators $L^a$
of a Lie algebra.
The extended non-Abelian gauge transformations of the
tensor gauge fields are defined
by the following equations  \cite{Savvidy:2005fi,Savvidy:2005zm,Savvidy:2005ki}:
\beqa\label{polygauge}
\delta A^{a}_{\mu} &=& ( \delta^{ab}\partial_{\mu}
+g f^{acb}A^{c}_{\mu})\xi^b ,~~~~~\\
\delta A^{a}_{\mu\nu} &=&  ( \delta^{ab}\partial_{\mu}
+  g f^{acb}A^{c}_{\mu})\xi^{b}_{\nu} + g f^{acb}A^{c}_{\mu\nu}\xi^{b},\nonumber\\
\delta A^{a}_{\mu\nu \lambda}& =&  ( \delta^{ab}\partial_{\mu}
+g f^{acb} A^{c}_{\mu})\xi^{b}_{\nu\lambda} +
g f^{acb}(  A^{c}_{\mu  \nu}\xi^{b}_{\lambda } +
A^{c}_{\mu \lambda }\xi^{b}_{ \nu}+
A^{c}_{\mu\nu\lambda}\xi^{b}),\nn\\
.........&.&............................\nn
\eeqa
where $\xi^{a}_{\lambda_1 ... \lambda_{s}}(x)$ are totally symmetric gauge parameters.
These extended gauge transformations
generate a closed algebraic structure.
The generalized field strengths  are defined as
\cite{Savvidy:2005fi,Savvidy:2005zm,Savvidy:2005ki}
\beqa\label{fieldstrengthparticular}
G^{a}_{\mu\nu} &=&
\partial_{\mu} A^{a}_{\nu} - \partial_{\nu} A^{a}_{\mu} +
g f^{abc}~A^{b}_{\mu}~A^{c}_{\nu},\\
G^{a}_{\mu\nu,\lambda} &=&
\partial_{\mu} A^{a}_{\nu\lambda} - \partial_{\nu} A^{a}_{\mu\lambda} +
g f^{abc}(~A^{b}_{\mu}~A^{c}_{\nu\lambda} + A^{b}_{\mu\lambda}~A^{c}_{\nu} ~),\nn\\
G^{a}_{\mu\nu,\lambda\rho} &=&
\partial_{\mu} A^{a}_{\nu\lambda\rho} - \partial_{\nu} A^{a}_{\mu\lambda\rho} +
g f^{abc}(~A^{b}_{\mu}~A^{c}_{\nu\lambda\rho} +
 A^{b}_{\mu\lambda}~A^{c}_{\nu\rho}+A^{b}_{\mu\rho}~A^{c}_{\nu\lambda}
 + A^{b}_{\mu\lambda\rho}~A^{c}_{\nu} ~),\nn\\
 ......&.&............................................\nn
\eeqa
and transform homogeneously with respect to the extended
gauge transformations (\ref{polygauge}). The field strength tensors are
antisymmetric in their first two indices and are totally symmetric with respect to the
rest of the indices.

These field strength tensors allow to
construct two series of gauge invariant quadratic forms. The
first series is given by the formula \cite{Savvidy:2005fi,Savvidy:2005zm,Savvidy:2005ki}
\beqa\label{fulllagrangian1}
{{\cal L}}_{s+1}&=&-{1\over 4} ~
G^{a}_{\mu\nu, \lambda_1 ... \lambda_s}~
G^{a}_{\mu\nu, \lambda_{1}...\lambda_{s}} +.......\nonumber\\
&=& -{1\over 4}\sum^{2s}_{i=0}~a^{s}_i ~
G^{a}_{\mu\nu, \lambda_1 ... \lambda_i}~
G^{a}_{\mu\nu, \lambda_{i+1}...\lambda_{2s}}
(\sum_{P} \eta^{\lambda_{i_1} \lambda_{i_2}} .......
\eta^{\lambda_{i_{2s-1}} \lambda_{i_{2s}}})~,
\eeqa
where the sum $\sum_P$ runs over all nonequal permutations of $i's$, in total $(2s-1)!!$
terms and the numerical coefficients are $a^{s}_i = {s!\over i!(2s-i)!}$.

The second series of gauge invariant quadratic forms is given by the formula
\cite{Savvidy:2005fi,Savvidy:2005zm,Savvidy:2005ki}:
\beqa\label{secondfulllagrangian}
{{\cal L}}^{'}_{s+1}&=&{1\over 4} ~
G^{a}_{\mu\nu,\rho\lambda_3  ... \lambda_{s+1}}~
G^{a}_{\mu\rho,\nu\lambda_{3} ...\lambda_{s+1}} +{1\over 4} ~
G^{a}_{\mu\nu,\nu\lambda_3  ... \lambda_{s+1}}~
G^{a}_{\mu\rho,\rho\lambda_{3} ...\lambda_{s+1}} +.......\nonumber\\
&=& {1\over 4}\sum^{2s+1}_{i=1}~{ a^{s}_{i-1}\over s}  ~
G^{a}_{\mu\lambda_1,\lambda_2  ... \lambda_i}~
G^{a}_{\mu\lambda_{2s+2},\lambda_{i+2} ...\lambda_{2s+1}}
(\sum^{'}_{P} \eta^{\lambda_{i_1} \lambda_{i_2}} .......
\eta^{\lambda_{i_{2s+1}} \lambda_{i_{2s+2}}})~,
\eeqa
where the sum $\sum^{'}_P$ runs over all nonequal permutations of $i's$, with exclusion
of the terms which contain $\eta^{\lambda_{1},\lambda_{2s+2}}$.
In order to make all tensor gauge fields dynamical one should add
the corresponding kinetic terms. Thus the invariant
Lagrangian describing dynamical tensor gauge bosons of all ranks
has the form
\be\label{fulllagrangian2}
{{\cal L}} = {{\cal L}}_{YM} + g_{2}{{\cal L}}_{2}+ g^{'}_{2}{{\cal L}}^{'}_{2}
+g_{3}{{\cal L}}_{3}+ g^{'}_{3}{{\cal L}}^{'}_{3}+...,
\ee
where ${{\cal L}}_{1} \equiv {{\cal L}}_{YM}$ and $g_{s}$ and $g^{'}_{s}~(s=2,3,...)$ are
arbitrary coupling constants.

Generally speaking, equations which follow from this
Lagrangian may contain propagating negative norm states.
Our main task in this article is to analyze particle spectrum of the
theory and to prove that
there are no negative modes in rank-2 and rank-3 tensor gauge fields in
four-dimensional space-time.
The problem is that tensor gauge fields have many components,
some of them define propagating physical modes of definite helicities,
but some of them may correspond to unphysical negative norm states.
As we shall see, only for specific values of the coupling constants
$g^{'}_{s}$ and only in four-dimensional space-time
the particle spectrum becomes completely physical.

Indeed, analyzing a linear sum of gauge invariant forms
\cite{Savvidy:2005fi,Savvidy:2005zm,Savvidy:2005ki}
$$
g_{2}{{\cal L}}_{2}+ g^{'}_{2}{{\cal L}}^{'}_{2}~,
$$
which are describing the propagation of the rank-2 tensor gauge
field $A^{a}_{\mu\lambda}$, we found
 that
if one chooses the coupling constant
$g^{'}_{2}=g_2$ then the sum
exhibits invariance with respect to
a bigger gauge group. In that case the free field equation
(\ref{freeequations1})/(\ref{mainequation})
for the rank-2 tensor gauge field
describes the propagation of {\it helicity-two and helicity-zero massless charged
tensor gauge bosons} and there are no propagating negative norm states.
Therefore the gauge invariant Lagrangian for the lower-rank tensor
gauge fields has the form \cite{Savvidy:2005fi,Savvidy:2005zm,Savvidy:2005ki}:
\beqa\label{totalactiontwo}
{{\cal L}}_2 +  {{\cal L}}^{'}_2  =
&-&{1\over 4}G^{a}_{\mu\nu,\lambda}G^{a}_{\mu\nu,\lambda}
-{1\over 4}G^{a}_{\mu\nu}G^{a}_{\mu\nu,\lambda\lambda}\\
&+&{1\over 4}G^{a}_{\mu\nu,\lambda}G^{a}_{\mu\lambda,\nu}
+{1\over 4}G^{a}_{\mu\nu,\nu}G^{a}_{\mu\lambda,\lambda}
+{1\over 2}G^{a}_{\mu\nu}G^{a}_{\mu\lambda,\nu\lambda}.\nn
\eeqa
Our aim here is to extend this analysis to the case of rank-3 tensor gauge
field $A_{\mu\lambda_1\lambda_2}$. The explicit form of $\CL_3$ and ${{\cal L}}^{'}_{3}$
 can be obtained from our general
formulas (\ref{fulllagrangian1}), (\ref{secondfulllagrangian}) and
(\ref{fulllagrangian2}) by substituting $s=2$. We shall consider the linear sum
$$
g_{3}{{\cal L}}_{3}+ g^{'}_{3}{{\cal L}}^{'}_{3}
$$
and demonstrate that for an appropriate choice of the coupling
constant $g^{'}_{3} =  {4\over 3} g_{3}$ the system
exhibits invariance with respect to
a bigger gauge group \cite{Savvidy:2006qf}.  The explicit description of this symmetry
together with the corresponding free field equation (\ref{freethirdrankequations})/
(\ref{freeequation})
for the rank-3 tensor gauge field  will be given
in the forth and fifth sections. We shall demonstrate that the free
equation (\ref{freethirdrankequations})/
(\ref{freeequation}) describes the
{\it propagation of helicity-three and a doublet of helicity-one
massless charged gauge bosons} and that
there are no propagating negative norm states.
Thus the Lagrangian for rank-3 non-Abelian gauge field will take the form
\beqa\label{thirdranktensorlagrangian}
{{\cal L}}_3 + {4 \over 3} {{\cal L}}^{'}_3
=&-&{1\over 4}G^{a}_{\mu\nu,\lambda\rho}G^{a}_{\mu\nu,\lambda\rho}
-{1\over 8}G^{a}_{\mu\nu ,\lambda\lambda}G^{a}_{\mu\nu ,\rho\rho}
-{1\over 2}G^{a}_{\mu\nu,\lambda}  G^{a}_{\mu\nu ,\lambda \rho\rho}
-{1\over 8}G^{a}_{\mu\nu}  G^{a}_{\mu\nu ,\lambda \lambda\rho\rho}+ \nn\\
&+&{1\over 3}G^{a}_{\mu\nu,\lambda\rho}G^{a}_{\mu\lambda,\nu\rho}+
{1\over 3} G^{a}_{\mu\nu,\nu\lambda}G^{a}_{\mu\rho,\rho\lambda}+
{1\over 3}G^{a}_{\mu\nu,\nu\lambda}G^{a}_{\mu\lambda,\rho\rho}+\\
&+&{1\over 3}G^{a}_{\mu\nu,\lambda}G^{a}_{\mu\lambda,\nu\rho\rho}
+{2\over 3}G^{a}_{\mu\nu,\lambda}G^{a}_{\mu\rho,\nu\lambda\rho}
+{1\over 3}G^{a}_{\mu\nu,\nu}G^{a}_{\mu\lambda,\lambda\rho\rho}
+{1\over 3}G^{a}_{\mu\nu}G^{a}_{\mu\lambda,\nu\lambda\rho\rho}.\nn
\eeqa

\section{\it Propagating Modes of Rank-2 Gauge Field}

Let us first consider the rank-2 gauge field $A^a_{\mu\lambda}$
\cite{Savvidy:2005fi,Savvidy:2005zm,Savvidy:2005ki}.
There are two invariant forms for the rank-2 tensor gauge
field ${{\cal L}}_2$ and ${{\cal L}}^{'}_2$ and we have to consider their
linear combination
$
g_2{{\cal L}}_2 + g^{'}_2 {{\cal L}}^{'}_2.
$
A free field equation of motion is defined by the quadratic part of this invariant, its cubic and quartic
parts define interaction.
The quadratic part of the ${{\cal L}}_2$ is
$$
{{\cal L}}^{quadratic}_2 =
{1 \over 2} A^{a}_{\alpha\acute{\alpha}}
H_{\alpha\acute{\alpha}\gamma\acute{\gamma}} A^{a}_{\gamma\acute{\gamma}} ,
$$
where the kinetic operator $H$ in momentum representation has the form
$$
H_{\alpha\acute{\alpha}\gamma\acute{\gamma}}(k)=
(-k^2 \eta_{\alpha\gamma} +k_{\alpha}k_{\gamma})
\eta_{\acute{\alpha}\acute{\gamma}}.
$$
It is obviously invariant with respect to the gauge
transformation $\delta A^{a}_{\mu\lambda} =\partial_{\mu} \xi^{a}_{\lambda}$,
but it is not invariant with respect to the alternative gauge transformations
$\tilde{\delta} A^{a}_{\mu \lambda} =\partial_{\lambda} \eta^{a}_{\mu}$. This can be
seen, for example, from the following relations in momentum representation:
\be\label{currentdivergence}
k_{\alpha}H_{\alpha\acute{\alpha}\gamma\acute{\gamma}}(k)=0,~~~
k_{\acute{\alpha}}H_{\alpha\acute{\alpha}\gamma\acute{\gamma}}(k)=
-(k^2 \eta_{\alpha\gamma} - k_{\alpha}k_{\gamma})k_{\acute{\gamma}} \neq 0 .
\ee
The quadratic part of the ${{\cal L}}^{'}_{2}$  is
\beqa
{{\cal L}}^{' quadratic}_{2}  =
{1 \over 2} A^{a}_{\alpha\acute{\alpha}}
H^{~'}_{\alpha\acute{\alpha}\gamma\acute{\gamma}} A^{a}_{\gamma\acute{\gamma}},
\eeqa
where the kinetic operator $H^{'}$ has the form
$$
H^{'}_{\alpha\acute{\alpha}\gamma\acute{\gamma}}(k)=
{1 \over 2}(\eta_{\alpha\acute{\gamma}}\eta_{\acute{\alpha}\gamma}
+\eta_{\alpha\acute{\alpha}}\eta_{\gamma\acute{\gamma}})k^2
-{1 \over 2}(\eta_{\alpha\acute{\gamma}}k_{\acute\alpha}k_{\gamma}
+\eta_{\acute\alpha\gamma}k_{\alpha}k_{\acute{\gamma}}
+\eta_{\alpha\acute\alpha}k_{\gamma}k_{\acute{\gamma}}
+\eta_{\gamma\acute{\gamma}}k_{\alpha}k_{\acute\alpha}
-2\eta_{\alpha\gamma}k_{\acute\alpha}k_{\acute{\gamma}}).
$$
It is also invariant with respect to the gauge
transformation $\delta A^{a}_{\mu\lambda} =\partial_{\mu} \xi^{a}_{\lambda}$,
but it is not invariant with respect to the gauge transformations
$\tilde{\delta} A^{a}_{\mu \lambda} =\partial_{\lambda} \eta^{a}_{\mu}$, as one can
see from analogous relations
\be\label{currentdivergenceprime}
k_{\alpha}H^{'}_{\alpha\acute{\alpha}\gamma\acute{\gamma}}(k)=0,~~~
k_{\acute{\alpha}}H^{'}_{\alpha\acute{\alpha}\gamma\acute{\gamma}}(k)=
(k^2 \eta_{\alpha\gamma} -k_{\alpha}k_{\gamma})k_{\acute{\gamma}} \neq 0 .
\ee
As it is obvious from (\ref{currentdivergence}) and
(\ref{currentdivergenceprime}), the sum
${{\cal L}}_2 + {{\cal L}}^{'}_2$, when $g^{'}_2=g_2$,  becomes invariant
with respect to the alternative gauge transformations
$\tilde{\delta} A^{a}_{\mu \lambda} =\partial_{\lambda} \eta^{a}_{\mu}$ and the kinetic
operator now has both of the symmetries:
\be\label{largegaugetransformation}
\delta A^{a}_{\mu \lambda} =\partial_{\mu} \xi^{a}_{\lambda}+
\partial_{\lambda} \eta^{a}_{\mu},
\ee
because\footnote{Longitudinal pieces in (\ref{currentdivergence}) and
(\ref{currentdivergenceprime}) cancel each other and the kinetic operator is
fully transversal.}
\be\label{zeroderivatives}
k_{\alpha}(H_{\alpha\acute{\alpha}\gamma\acute{\gamma}}+
H^{'}_{\alpha\acute{\alpha}\gamma\acute{\gamma}})=0,~~~
k_{\acute{\alpha}}(H_{\alpha\acute{\alpha}\gamma\acute{\gamma}}+
H^{'}_{\alpha\acute{\alpha}\gamma\acute{\gamma}})=0 .
\ee
Thus our kinetic operator is a sum
\beqa\label{totalfreelagrangian}
{{\cal L}}_2 +   {{\cal L}}^{'}_2 ~\vert_{quadratic}={1 \over 2} A^{a}_{\alpha\acute{\alpha}}
(H_{\alpha\acute{\alpha}\gamma\acute{\gamma}}+
H^{'}_{\alpha\acute{\alpha}\gamma\acute{\gamma}}) A^{a}_{\gamma\acute{\gamma}}
={1 \over 2} A^{a}_{\alpha\acute{\alpha}}
\CH_{\alpha\acute{\alpha}\gamma\acute{\gamma}}  A^{a}_{\gamma\acute{\gamma}},
\eeqa
where
\beqa\label{quadraticform}
\CH_{\alpha\acute{\alpha}\gamma\acute{\gamma}}(k)=
(-\eta_{\alpha\gamma}\eta_{\acute{\alpha}\acute{\gamma}}
+{1 \over 2}\eta_{\alpha\acute{\gamma}}\eta_{\acute{\alpha}\gamma}
+{1 \over 2}\eta_{\alpha\acute{\alpha}}\eta_{\gamma\acute{\gamma}})k^2
+\eta_{\alpha\gamma}k_{\acute\alpha}k_{\acute{\gamma}}
+\eta_{\acute\alpha \acute{\gamma}}k_{\alpha}k_{\gamma}\nn\\
-{1 \over 2}(\eta_{\alpha\acute{\gamma}}k_{\acute\alpha}k_{\gamma}
+\eta_{\acute\alpha\gamma}k_{\alpha}k_{\acute{\gamma}}
+\eta_{\alpha\acute\alpha}k_{\gamma}k_{\acute{\gamma}}
+\eta_{\gamma\acute{\gamma}}k_{\alpha}k_{\acute\alpha}).
\eeqa
In terms of field strength tensor the quadratic part is
\be
\label{kineticterm}
{{\cal L}}_2 +   {{\cal L}}^{'}_2 ~\vert_{quadratic}  = -{1\over
4}F^{a}_{\mu\nu,\lambda}F^{a}_{\mu\nu,\lambda} +
 {1\over 4}F^{a}_{\mu\nu,\lambda}F^{a}_{\mu\lambda,\nu}+{1\over
4}F^{a}_{\mu\nu,\nu}F^{a}_{\mu\lambda,\lambda},
\ee
where
\be\label{freefieldstrength}
F^{a}_{\mu\nu,\lambda}=
\partial_{\mu} A^{a}_{\nu\lambda} - \partial_{\nu} A^{a}_{\mu\lambda},
\ee
and the equation of motion takes the form
\beqa\label{freeequations1}
\partial^{\mu} F^{a}_{\mu\nu,\lambda}
-{1\over 2} (\partial^{\mu} F^{a}_{\mu\lambda,\nu}
+\partial^{\mu} F^{a}_{\lambda\nu,\mu}
+\partial_{\lambda}F^{a~~\mu}_{\mu\nu,}
+\eta_{\nu\lambda} \partial^{\mu}F^{a~~\rho}_{\mu\rho,}) = 0.
\eeqa
In terms of tensor gauge field
the free equation of motion (\ref{freeequations1}) is
\beqa\label{mainequation}
\partial^{2}(A^{a}_{\nu\lambda} -{1\over 2}A^{a}_{\lambda\nu})
-\partial_{\nu} \partial_{\mu}  (A^{a}_{\mu\lambda}-
{1\over 2}A^{a}_{\lambda\mu} )&-&
\partial_{\lambda} \partial_{\mu}  (A^{a}_{\nu\mu} - {1\over 2}A^{a}_{\mu\nu} )
+\partial_{\nu} \partial_{\lambda} ( A^{a}_{\mu\mu}-{1\over 2}A^{a}_{\mu\mu})\nn\\
&+&{1\over 2}\eta_{\nu\lambda} ( \partial_{\mu} \partial_{\rho}A^{a}_{\mu\rho}
-  \partial^{2}A^{a}_{\mu\mu})=0
\eeqa
and it describes the propagation of massless charged gauge bosons
of helicity two and zero. Indeed, this can be seen by decomposition of the rank-2 gauge
field into symmetric and antisymmetric parts. For the symmetric
tensor gauge fields
$A^{a}_{\nu\lambda} = A^{a}_{\lambda\nu}$ our equation
reduces to the  Einstein and Fierz-Pauli equation
$$
\partial^{2} A_{\nu\lambda}
-\partial_{\nu} \partial_{\mu}  A_{\mu\lambda} -
\partial_{\lambda} \partial_{\mu}  A_{\mu\nu}
+ \partial_{\nu} \partial_{\lambda}  A_{\mu\mu}
+\eta_{\nu\lambda}  (\partial_{\mu} \partial_{\rho}A_{\mu\rho}
- \partial^{2} A_{\mu\mu}) =0,
$$
which describes the propagation of massless gauge boson of helicity two.
For the antisymmetric fields it reduces to the Kalb-Ramond equation
$$
\partial^{2} A_{\nu\lambda}
-\partial_{\nu} \partial_{\mu}  A_{\mu\lambda} +
\partial_{\lambda} \partial_{\mu}  A_{\mu\nu} =0
$$
and describes the propagation of helicity-zero state.

A more direct way to solve the free equation of motion (\ref{freeequations1})/(\ref{mainequation})
is to consider it in the momentum representation \cite{Konitopoulos:2007hw}:
\be\label{basicequation}
\CH_{\alpha\acute{\alpha} \gamma\acute{\gamma}}(k)~
A^{a}_{\gamma\acute{\gamma}} =0.
\ee
The vector space of independent solutions
$A_{\gamma\acute{\gamma}}$ crucially depends on the rank  of the matrix
$H_{\alpha\acute{\alpha} \gamma\acute{\gamma}}(k)$. If the matrix
operator $H$ has dimension $d \times d$ and its rank  is $rankH=r$,
then the vector space of solutions has the dimension
$$
{\cal N}= d-r.
$$
Because the matrix operator $H_{\alpha\acute{\alpha} \gamma\acute{\gamma}}(k)$
explicitly depends on the momentum $k_{\mu}$, its $rankH=r$ also depends on
momenta and therefore the number of independent solutions
${\cal N}$  depends on momenta
\be
{\cal N}(k) = d-r(k)~.
\ee
Analyzing the $rankH$ of the matrix operator $H$
one can observe that it depends on the value of momentum
square $k_{\mu}^2$. When $k_{\mu}^2 \neq 0$ - off mass-shell momenta - the
vector space consists of {\it pure gauge fields}. When $k_{\mu}^2 = 0$ -
on mass-shell
momenta - the vector space consists of {\it pure gauge fields and propagating
modes}. Therefore the number of propagating modes can be calculated from
the following relation:
\be\label{numberofpropagationmodes}
\sharp ~ of~ propagating ~modes= {\cal N}(k)\vert_{k^2 = 0} -
{\cal N}(k)\vert_{k^2 \neq 0}=
 rankH \vert_{k^2\ne 0}-rankH \vert_{k^2=0}.
\ee

Our field equation (\ref{basicequation})
for the tensor gauge field $A_{\mu\lambda}$ is defined
by the matrix operator (\ref{quadraticform}), which
in the four-dimensional space-time is a $16 \times 16$ matrix\footnote{The multi-index $A \equiv
(\mu,\lambda)$ takes sixteen values.}.
In the reference frame, where
$k^{\gamma}=(\omega,0,0,k)$, it has a particularly simple form.
If $\omega^2 - k^2 \neq 0$, the rank of the 16-dimensional
matrix
$
H_{\alpha\acute{\alpha}\gamma\acute{\gamma}}(k)
$
is equal to $rankH\vert_{\omega^2 - k^2 \ne 0}=9$
and the number of linearly independent solutions is $16-9=7$.
These seven solutions are
\beqa\label{explicitgaugesolutions}
 e_{\gamma\acute{\gamma}}=
\left(\begin{array}{cccc}
-\omega^{2}&0&0&0 \\
0&0&0&0 \\
0&0&0&0 \\
0&0&0&k^{2} \\
\end{array} \right)
&,&\left(\begin{array}{cccc}
\omega&0&0&0 \\
0&0&0&0 \\
0&0&0&0 \\
k&0&0&0 \\
\end{array} \right),
\left(\begin{array}{cccc}
0&\omega&0&0 \\
0&0&0&0 \\
0&0&0&0 \\
0&k&0&0 \\
\end{array} \right),\left(\begin{array}{cccc}
0&0&\omega&0 \\
0&0&0&0 \\
0&0&0&0 \\
0&0&k&0 \\
\end{array} \right),\nn\\ \left(\begin{array}{cccc}
\omega&0&0&k \\
0&0&0&0 \\
0&0&0&0 \\
0&0&0&0 \\
\end{array} \right)&,&\left(\begin{array}{cccc}
0&0&0&0 \\
\omega&0&0&k \\
0&0&0&0 \\
0&0&0&0 \\
\end{array} \right),\left(\begin{array}{cccc}
0&0&0&0 \\
0&0&0&0 \\
\omega&0&0&k \\
0&0&0&0 \\
\end{array} \right)
\eeqa
and they are pure gauge fields
\be\label{puregaugepotentials}
e_{\gamma\acute{\gamma}}=
k_{\gamma}\xi_{\acute{\gamma}}+k_{\acute{\gamma}}\eta_{\gamma}.
\ee
When $\omega^2 - k^2 = 0$, then the rank of the matrix
$
H_{\alpha\acute{\alpha}\gamma\acute{\gamma}}(k)
$
drops and is equal to $rankH\vert_{\omega^2 - k^2 = 0}  =6$.
This leaves us with $16-6=10$ solutions. These are 7 solutions,
the pure gauge potentials (\ref{explicitgaugesolutions}), (\ref{puregaugepotentials}),
and three new solutions representing the propagating modes:
\beqa\label{physicalmodes}
e_{\gamma\acute{\gamma}}^{(1)}=\left(\begin{array}{cccc}
0&0&0&0 \\
0&-1&0&0 \\
0&0&1&0 \\
0&0&0&0 \\
\end{array} \right),~~
e_{\gamma\acute{\gamma}}^{(2)}=
\left(\begin{array}{cccc}
0&0&0&0 \\
0&0&1&0 \\
0&1&0&0 \\
0&0&0&0 \\
\end{array} \right),~~
e_{\gamma\acute{\gamma}}^{A}=
\left(\begin{array}{cccc}
0&0&0&0 \\
0&0&1&0 \\
0&-1&0&0 \\
0&0&0&0 \\
\end{array} \right)
\eeqa
Thus the
general solution of the equation on the mass-shell is
\be\label{gensolution}
e_{\gamma\acute{\gamma}}=\xi_{\acute{\gamma}}k_{\gamma} +
\eta_{\gamma} k_{\acute{\gamma}}+
c_{1}e^{(1)}_{\gamma\acute{\gamma}}+c_{2}e^{(2)}_{\gamma\acute{\gamma}}
+c_3 e^{(A)}_{\gamma\acute{\gamma}},
\ee
where $c_{1},c_{2}, c_{3}$ are arbitrary constants.
We see that the number of the propagating modes  is three:
$$
rankH\vert_{\omega^2 - k^2 \ne 0}-
rankH\vert_{\omega^2 - k^2 = 0}=9-6=3.
$$
These are the propagating modes of {\it helicity-two and helicity-zero
$\lambda = \pm 2, 0$ charged gauge bosons}
\cite{Savvidy:2005fi,Savvidy:2005zm,Savvidy:2005ki}.

The above consideration brings the final form of the gauge invariant Lagrangian
for the rank-2 tensor gauge field to the form (\ref{totalactiontwo}) with its
free equation of motion (\ref{freeequations1})/(\ref{mainequation}). And, as we
have seen, has a well defined physical spectrum (\ref{gensolution}).

\section{\it Rank-3 Tensor Gauge Field}

Let us turn now to the rank-3 gauge field.
There are two invariant forms ${{\cal L}}_3$ and ${{\cal L}}^{'}_3$
for the rank-3 tensor gauge
field  $A^{a}_{\mu\nu\lambda}$ and we have to consider their
linear combination
$
g_3{{\cal L}}_3 + g^{'}_3 {{\cal L}}^{'}_3
$.
The Lagrangian ${{\cal L}}_3$ has
the form (\ref{fulllagrangian1})\footnote{In (\ref{fulllagrangian1})
one should take s=2.}
\cite{Savvidy:2005fi,Savvidy:2005zm,Savvidy:2005ki}
\beqa
{{\cal L}}_3 =-{1\over 4}G^{a}_{\mu\nu,\lambda\rho}G^{a}_{\mu\nu,\lambda\rho}
-{1\over 8}G^{a}_{\mu\nu ,\lambda\lambda}G^{a}_{\mu\nu ,\rho\rho}
-{1\over 2}G^{a}_{\mu\nu,\lambda}  G^{a}_{\mu\nu ,\lambda \rho\rho}
-{1\over 8}G^{a}_{\mu\nu}  G^{a}_{\mu\nu ,\lambda \lambda\rho\rho}~,
\eeqa
where higher rank field strength tensors are:
\beqa\label{spin4fieldstrenghth}
G^{a}_{\mu\nu ,\lambda \rho \sigma} =
\partial_{\mu} A^{a}_{\nu \lambda \rho \sigma} -
\partial_{\nu} A^{a}_{\mu \lambda\rho\sigma} +
g f^{abc}\{~A^{b}_{\mu}~A^{c}_{\nu \lambda \rho\sigma}
+A^{b}_{\mu\lambda}~A^{c}_{\nu\rho \sigma} +
A^{b}_{\mu\rho }~A^{c}_{\nu\lambda\sigma} +
A^{b}_{\mu\sigma}~A^{c}_{\nu\lambda\rho} +\nn\\
+A^{b}_{\mu\lambda\rho}~A^{c}_{\nu \sigma} +
A^{b}_{\mu\lambda\sigma}~A^{c}_{\nu\rho} +
A^{b}_{\mu\rho\sigma}~A^{c}_{\nu \lambda} +
     A^{b}_{\mu\lambda\rho\sigma }~A^{c}_{\nu} ~\}\nonumber
\eeqa
and
\beqa\label{spin4fieldstrenghth4}
G^{a}_{\mu\nu ,\lambda \rho \sigma\delta} =
\partial_{\mu} A^{a}_{\nu \lambda \rho \sigma\delta} -
\partial_{\nu} A^{a}_{\mu \lambda\rho\sigma\delta} &+&
g f^{abc}\{~A^{b}_{\mu}~A^{c}_{\nu \lambda \rho\sigma\delta}
+\sum_{ \lambda \leftrightarrow \rho,\sigma,\delta}
        A^{b}_{\mu\lambda}~A^{c}_{\nu\rho \sigma\delta} + \nn\\
&+&\sum_{\lambda,\rho \leftrightarrow \sigma,\delta}
        A^{b}_{\mu\lambda\rho}~A^{c}_{\nu\sigma\delta} +
\sum_{\lambda,\rho,\sigma\leftrightarrow \delta}
       A^{b}_{\mu\lambda\rho\sigma}~A^{c}_{\nu\delta} +
     A^{b}_{\mu\lambda\rho\sigma\delta }~A^{c}_{\nu} ~\}.\nonumber
\eeqa
The terms in parentheses are symmetric over $\lambda \rho\sigma$ and
$\lambda \rho \sigma\delta$ respectively. The Lagrangian ${{\cal L}}_3$
is invariant with respect to the extended gauge transformations  (\ref{polygauge})
of the low-rank gauge fields
$ A_{\mu}, A_{\mu\nu}, A_{\mu\nu\lambda}$  together with the fourth- and fifth-rank gauge fields
\beqa\label{gaugetransform4}
\delta_{\xi}  A_{\mu\nu\lambda\rho} &=&\partial_{\mu}\xi_{\nu\lambda\rho}
-i g[A_{\mu},\xi_{\nu\lambda\rho}]
-i g [A_{\mu\nu},\xi_{\lambda\rho}]
-i g [A_{\mu\lambda},\xi_{\nu\rho}]
-i g [A_{\mu\rho},\xi_{\nu\lambda}]-\nn\\
&-&i g  [A_{\mu\nu\lambda},\xi_{\rho}]
-i g  [A_{\mu\nu\rho},\xi_{\lambda}]
-i g  [A_{\mu\lambda\rho},\xi_{\nu}]
-i g [A_{\mu\nu\lambda\rho},\xi]\nonumber,\\
\delta_{\xi}  A_{\mu\nu\lambda\rho\sigma} &=&\partial_{\mu}\xi_{\nu\lambda\rho\sigma}
-i g[A_{\mu},\xi_{\nu\lambda\rho\sigma}]
-i g \sum_{\nu \leftrightarrow \lambda\rho\sigma}
[A_{\mu\nu},\xi_{\lambda\rho\sigma}]-\nn\\
   &-&ig\sum_{\nu\lambda \leftrightarrow \rho\sigma}
   [A_{\mu\nu\lambda},\xi_{\rho\sigma}]
      -ig\sum_{\nu\lambda\rho \leftrightarrow \sigma}
      [A_{\mu\nu\lambda\rho},\xi_{\sigma}]
-i g [A_{\mu\nu\lambda\rho},\xi], ~\nonumber
\eeqa
where the gauge parameters $\xi_{\nu\lambda\rho}$ and $\xi_{\nu\lambda\rho\sigma}$
are totally symmetric  tensors.
The second Lagrangian ${{\cal L}}^{'}_3$ has
the form (\ref{secondfulllagrangian})\footnote{In (\ref{secondfulllagrangian})
one should take s=2.} \cite{Savvidy:2005fi,Savvidy:2005zm,Savvidy:2005ki}
\beqa\label{actionthreeprime}
{{\cal L}}^{'}_3 &=&  {1\over 4}
G^{a}_{\mu\nu,\lambda\rho}G^{a}_{\mu\lambda,\nu\rho}+
{1\over 4} G^{a}_{\mu\nu,\nu\lambda}G^{a}_{\mu\rho,\rho\lambda}+
{1\over 4}G^{a}_{\mu\nu,\nu\lambda}G^{a}_{\mu\lambda,\rho\rho}\nn\\
&+&{1\over 4}G^{a}_{\mu\nu,\lambda}G^{a}_{\mu\lambda,\nu\rho\rho}
+{1\over 2}G^{a}_{\mu\nu,\lambda}G^{a}_{\mu\rho,\nu\lambda\rho}
+{1\over 4}G^{a}_{\mu\nu,\nu}G^{a}_{\mu\lambda,\lambda\rho\rho}
+{1\over 4}G^{a}_{\mu\nu}G^{a}_{\mu\lambda,\nu\lambda\rho\rho}.
\eeqa
We wish to know
if there exists a special value of the constant $g^{'}_3$ at which
the system will have higher symmetry, as it happens in the case
of the rank-2 gauge field. We shall see that this indeed takes place.

A free field equation of motion is defined by the quadratic part of invariant
$
g_3{{\cal L}}_3 + g^{'}_3 {{\cal L}}^{'}_3
$, the interaction - by cubic and quartic.
The quadratic part of the Lagrangian ${{\cal L}}_3$ comes from the terms
\beqa
-{1\over 4}G^{a}_{\mu\nu,\lambda\rho}G^{a}_{\mu\nu,\lambda\rho}
-{1\over 8}G^{a}_{\mu\nu ,\lambda\lambda}G^{a}_{\mu\nu ,\rho\rho}
\eeqa
and has the form
$$
{{\cal L}}^{quadratic}_3
={1 \over 2} A^{a}_{\alpha\alpha^{'}\alpha^{''}}
H_{\alpha\alpha^{'}\alpha^{''}\gamma\gamma^{'}\gamma^{''}}
A^{a}_{\gamma\gamma^{'}\gamma^{''}}  ,
$$
where the kinetic operator $H$ in the momentum representation is
\beqa\label{rank3form1}
H_{\alpha\alpha^{'}\alpha^{''}\gamma\gamma^{'}\gamma^{''}}(k)=-{1 \over 2}
H_{\alpha\gamma}
(\eta_{\alpha^{'} \gamma^{'}} \eta_{ \alpha^{''}  \gamma^{''} } +
\eta_{\alpha^{'}\gamma^{''}} \eta_{\alpha^{''} \gamma^{'}}+
\eta_{\alpha^{'}\alpha^{''}} \eta_{\gamma^{'} \gamma^{''}}),
\eeqa
and $H_{\alpha\gamma}= k^2 \eta_{\alpha\gamma} - k_{\alpha}k_{\gamma}$.
It is invariant with respect to the gauge
transformation
$
\delta A^{a}_{\mu\nu\lambda} =\partial_{\mu} \xi^{a}_{\nu\lambda},
$
which can be seen from the relation
$
k_{\alpha}H_{\alpha\alpha^{'}\alpha^{''}\gamma\gamma^{'}\gamma^{''}}(k)=0.
$
But it is not invariant with respect to the alternative gauge transformations
$
\tilde{\delta} A^{a}_{\mu \nu\lambda} =\partial_{\nu} \zeta^{a}_{\mu\lambda}+
\partial_{\lambda} \zeta^{a}_{\mu\nu},
$
where the gauge parameter $\zeta^{a}_{\mu\lambda}$ is a
totally symmetric tensor.
This can be seen from the following relation in momentum representation
\beqa\label{firstlagrangiandivergence}
k_{\alpha^{'}}H_{\alpha\alpha^{'}\alpha^{''}\gamma\gamma^{'}\gamma^{''}}(k)&=&
-{1\over 2}H_{\alpha\gamma}~
(k_{\gamma^{'}} \eta_{ \alpha^{''}  \gamma^{''} } +
k_{\gamma^{''}} \eta_{\alpha^{''} \gamma^{'}}+
k_{\alpha^{''}} \eta_{\gamma^{'} \gamma^{''}}) \neq 0.
\eeqa
The quadratic part of the Lagrangian ${{\cal L}}^{'}_3$ comes from the terms
\beqa\label{freeactionthreeprime}
 {1\over 4}
G^{a}_{\mu\nu,\lambda\rho}G^{a}_{\mu\lambda,\nu\rho}+
{1\over 4} G^{a}_{\mu\nu,\nu\lambda}G^{a}_{\mu\rho,\rho\lambda}+
{1\over 4}G^{a}_{\mu\nu,\nu\lambda}G^{a}_{\mu\lambda,\rho\rho},
\eeqa
and has the form
$$
{{\cal L}}^{'~quadratic}_3
={1 \over 2} A^{a}_{\alpha\alpha^{'}\alpha^{''}}
H^{~'}_{\alpha\alpha^{'}\alpha^{''}\gamma\gamma^{'}\gamma^{''}}
A^{a}_{\gamma\gamma^{'}\gamma^{''}}  ,
$$
The kinetic operator $H^{'}$ is
(see Appendix A for derivation)
\beqa\label{rank3form2}
H^{~'}_{\alpha\alpha^{'}\alpha^{''}\gamma\gamma^{'}\gamma^{''}}(k)=
{1\over 8}\{ &+& (k^2 \eta_{\alpha\alpha^{'}}-k_{\alpha}k_{\alpha^{'}})
(\eta_{\alpha^{''}\gamma} \eta_{\gamma^{'} \gamma^{''}}
+\eta_{\alpha^{''}\gamma^{'}} \eta_{\gamma \gamma^{''}}
+\eta_{\alpha^{''}\gamma^{''}} \eta_{\gamma\gamma^{'} })\nn\\
&+& (k^2 \eta_{\alpha\alpha^{''}} -k_{\alpha}k_{\alpha^{''}})
(\eta_{\alpha^{'}\gamma} \eta_{\gamma^{'} \gamma^{''}}
+\eta_{\alpha^{'}\gamma^{'}} \eta_{\gamma \gamma^{''}}
+\eta_{\alpha^{'}\gamma^{''}} \eta_{\gamma\gamma^{'} })\nn\\
&+&(k^2 \eta_{\alpha\gamma^{'}}-k_{\alpha}k_{\gamma^{'}})
(\eta_{\alpha^{'}\gamma} \eta_{\alpha^{''} \gamma^{''}}
+\eta_{\alpha^{'}\gamma^{''}} \eta_{\alpha^{''}\gamma }
+\eta_{\alpha^{'}\alpha^{''}} \eta_{\gamma\gamma^{''} })\nn\\
&+&(k^2 \eta_{\alpha\gamma^{''}}-k_{\alpha}k_{\gamma^{''}})
(\eta_{\alpha^{'}\gamma} \eta_{\alpha^{''} \gamma^{'}}
+\eta_{\alpha^{'}\gamma^{'}} \eta_{\alpha^{''}\gamma }+
\eta_{\alpha^{'}\alpha^{''}} \eta_{\gamma \gamma^{'}})~\}\nn\\
             -{ 1\over 8}\{&+&k_{\gamma}k_{\alpha^{'}}
(\eta_{\alpha\gamma^{'}} \eta_{\alpha^{''} \gamma^{''}}
+\eta_{\alpha\gamma^{''}} \eta_{\alpha^{''} \gamma^{'}}
+\eta_{\alpha\alpha^{''}} \eta_{\gamma^{'}\gamma^{''} })\nn\\
&+&k_{\gamma}k_{\alpha^{''}}
(\eta_{\alpha\gamma^{'}} \eta_{\alpha^{'} \gamma^{''}}
+\eta_{\alpha\gamma^{''}} \eta_{\alpha^{'}\gamma^{'} }
+\eta_{\alpha\alpha^{'}} \eta_{\gamma^{'}\gamma^{''} })\nn\\
&+&k_{\gamma}k_{\gamma^{'}}
(\eta_{\alpha\alpha^{'}} \eta_{\alpha^{''} \gamma^{''}}
+\eta_{\alpha\alpha^{''}} \eta_{\alpha^{'} \gamma^{''}}
+\eta_{\alpha\gamma^{''}} \eta_{\alpha^{'}\alpha^{''} })\nn\\
&+&k_{\gamma}k_{\gamma^{''}}
(\eta_{\alpha\alpha^{'}} \eta_{\alpha^{''} \gamma^{'}}
+\eta_{\alpha\alpha^{''}} \eta_{\alpha^{'}\gamma^{'} }
+\eta_{\alpha\gamma^{'}} \eta_{\alpha^{'}\alpha^{''} })~\}\nn\\
   +{ 1\over 4}\{&+&\eta_{\alpha\gamma} (k_{\alpha^{'}}k_{\gamma^{'}}
\eta_{\alpha^{''} \gamma^{''}} + k_{\alpha^{'}}k_{\gamma^{''}}
\eta_{\alpha^{''} \gamma^{'}} + k_{\alpha^{''}}k_{\gamma^{'}}
\eta_{\alpha^{'} \gamma^{''}} \nn\\
&+&k_{\alpha^{''}}k_{\gamma^{''}}
\eta_{\alpha^{'} \gamma^{'}} + k_{\alpha^{'}}k_{\alpha^{''}}
\eta_{ \gamma^{'}\gamma^{''}} +k_{\gamma^{'}}k_{\gamma^{''}}
\eta_{\alpha^{'}\alpha^{''} })~\}.
\eeqa
It is again invariant with respect  to the transformation $\delta A^{a}_{\mu\nu\lambda} =
\partial_{\mu} \xi^{a}_{\nu\lambda}$ which is translated into the relation
$
k_{\alpha}H^{~'}_{\alpha\alpha^{'}\alpha^{''}\gamma\gamma^{'}\gamma^{''}}(k)=0,
$
but it is not invariant with respect to the transformation
$\tilde{\delta} A^{a}_{\mu \nu\lambda} =\partial_{\nu} \zeta^{a}_{\mu\lambda}
+\partial_{\lambda} \zeta^{a}_{\mu\nu}$,
as one can see from the relation (see also Appendix A for derivation)
\beqa\label{secondlagrangiandivergence}
k_{\alpha^{'}}H^{~'}_{\alpha\alpha^{'}\alpha^{''}\gamma\gamma^{'}\gamma^{''}}(k)=
{1\over 8}\{&+&H_{\alpha\alpha^{''}}
(k_{\gamma^{'}} \eta_{\gamma \gamma^{''}}
+k_{\gamma^{''}} \eta_{\gamma\gamma^{'} })\nn\\
&+&H_{\alpha\gamma^{'}}
(k_{\gamma^{''}} \eta_{\alpha^{''}\gamma }
+k_{\alpha^{''}} \eta_{\gamma\gamma^{''} })\nn\\
&+&H_{\alpha\gamma^{''}}
(k_{\gamma^{'}} \eta_{\alpha^{''}\gamma }+
k_{\alpha^{''}} \eta_{\gamma \gamma^{'}})~\}\nn\\
             -{ 1\over 4}\{
&+&k_{\gamma}k_{\alpha^{''}}
(k_{\gamma^{''}}\eta_{\alpha\gamma^{'}}
+k_{\gamma^{'} }\eta_{\alpha\gamma^{''}})
+k_{\gamma}k_{\gamma^{'}}
k_{\gamma^{''}}\eta_{\alpha\alpha^{''}}
-3\eta_{\alpha\gamma}k_{\alpha^{''}}k_{\gamma^{'}}k_{\gamma^{''}} ~\}\nn\\
   +{ 1\over 4}\{&+&H_{\alpha\gamma}
   (k_{\gamma^{'}}\eta_{\alpha^{''} \gamma^{''}}+
   k_{\gamma^{''}}\eta_{\alpha^{''} \gamma^{'}}+ k_{\alpha^{''}}
\eta_{ \gamma^{'}\gamma^{''}})~\} \neq 0.
\eeqa
We have to see now whether the total longitudinal part of the kinetic operator
$$
k_{\alpha^{'}} (g_3~H^{~'}_{\alpha\alpha^{'}\alpha^{''}\gamma\gamma^{'}\gamma^{''}}+
g^{'}_3~ H^{~'}_{\alpha\alpha^{'}\alpha^{''}\gamma\gamma^{'}\gamma^{''}}~)
$$
can be made equal to zero by an appropriate choice of the coupling constant $g^{'}_3$.
For that let us compare the expressions (\ref{firstlagrangiandivergence}) and
(\ref{secondlagrangiandivergence}) for longitudinal terms. As one can see, only the last term in
(\ref{secondlagrangiandivergence})
$
H_{\alpha\gamma}
   (k_{\gamma^{'}}\eta_{\alpha^{''} \gamma^{''}}+
   k_{\gamma^{''}}\eta_{\alpha^{''} \gamma^{'}}+ k_{\alpha^{''}}
\eta_{ \gamma^{'}\gamma^{''}})
$
and the whole term (\ref{firstlagrangiandivergence}) can  cancel each other if
we choose $g^{'}_3=2 g_3$, but this will leave the rest of the terms in (\ref{secondlagrangiandivergence})
untouched, thus
$
k_{\alpha^{'}} (g_3~H_{\alpha\alpha^{'}\alpha^{''}\gamma\gamma^{'}\gamma^{''}}+
g^{'}_3 H^{'}_{\alpha\alpha^{'}\alpha^{''}\gamma\gamma^{'}\gamma^{''}}~) \neq 0
$
for all values of $g^{'}_3$. This situation differs from the case of
the rank-2 gauge field. In the last case we were able
to choose coupling constant $g^{'}_2$ so that longitudinal pieces (\ref{currentdivergence}) and
(\ref{currentdivergenceprime}) cancel each other.

In order to understand the reason, why in the case of the rank-3 gauge field
it is impossible to fully cancel longitudinal pieces,
we have to remind a beautiful result obtained long ago by Schwinger \cite{schwinger}.
It has been proven by Schwinger
\cite{schwinger,Savvidy:2006qf,Guttenberg:2008qe}
that it is impossible to derive free
field equation for the totally symmetric rank-3 tensor which is invariant
with respect to the gauge group of  transformations
$\delta A_{\mu \nu\lambda} =\partial_{\mu} \xi^{a}_{\nu\lambda}+
\partial_{\nu} \xi_{\mu\lambda}+ \partial_{\lambda} \xi_{\mu\nu}$
without imposing some restriction on the gauge parameters $\xi_{\mu\nu}$.
As Schwinger demonstrated, the gauge parameter should be traceless: $\xi_{\mu\mu}=0$.
We shall see that similar phenomena take place also in our case, that is,
the gauge parameter $\zeta^{a}_{\mu\lambda}$ should
fulfill the restriction (\ref{restrictionongaugeparameters}).

What we would like to prove is
that our equation has enhanced invariance
with respect to the gauge group of transformations
\be\label{enhancedgaugetransformationrank3}
\tilde{\delta} A^{a}_{\mu \nu\lambda} =\partial_{\nu} \zeta^{a}_{\mu\lambda}+
\partial_{\lambda} \zeta^{a}_{\mu\nu},
\ee
only if the gauge parameter $\zeta^{a}_{\mu\lambda}$ fulfills the following
restriction:
\be\label{restrictionongaugeparameters}
\partial_{\rho}\zeta^{a}_{\rho\lambda}-\partial_{\lambda} \zeta^{a}_{ \rho\rho}=0.
\ee
This takes place when we choose
$
g^{'}_{3} =  {4\over 3} g_{3}~.
$
Indeed, let us consider the equation of motion.
From  ${{\cal L}}^{quadratic}_3$ we have:
\beqa
H_{\alpha\alpha^{'}\alpha^{''}\gamma\gamma^{'}\gamma^{''}}
A^{a}_{\gamma\gamma^{'}\gamma^{''}}
=\partial^{2} A^{a}_{\alpha\alpha^{'}\alpha^{''}} -\partial_{\alpha} \partial_{\rho}
A^{a}_{\rho\alpha^{'}\alpha^{''}} +
{1 \over 2}\eta_{\alpha^{'}\alpha^{''}}( \partial^{2} A^{a}_{\alpha\rho\rho}
-\partial_{\alpha} \partial_{\rho} A^{a}_{\rho\lambda\lambda}),
\eeqa
and from ${{\cal L}}^{'~quadratic}_3$
\beqa
H^{'}_{\alpha\alpha^{'}\alpha^{''}\gamma\gamma^{'}\gamma^{''}}
A^{a}_{\gamma\gamma^{'}\gamma^{''}}=-{1 \over 8}\{ \partial^{2}( A^{a}_{\alpha^{'}\alpha \alpha^{''}}+
A^{a}_{\alpha^{'}\alpha^{''}\alpha }+A^{a}_{\alpha^{''}\alpha \alpha^{'}}+
A^{a}_{\alpha^{''}\alpha^{'}\alpha })-\nn\\
-\partial_{\alpha} \partial_{\rho}(A^{a}_{\alpha^{'}\rho\alpha^{''}}
+A^{a}_{\alpha^{'}\alpha^{''}\rho}
+A^{a}_{\alpha^{''}\rho\alpha^{'}}
+A^{a}_{\alpha^{''}\alpha^{'}\rho})-\nn\\
-\partial_{\alpha^{'}} \partial_{\rho}(A^{a}_{\rho\alpha\alpha^{''}}
+A^{a}_{\rho\alpha^{''}\alpha} - A^{a}_{\alpha\rho\alpha^{''}}
-A^{a}_{\alpha\alpha^{''}\rho}-A^{a}_{\alpha\alpha^{''}\rho}
-A^{a}_{\alpha\rho\alpha^{''}})-\nn\\
-\partial_{\alpha^{''}} \partial_{\rho}(A^{a}_{\rho\alpha\alpha^{'}}
+A^{a}_{\rho\alpha^{'}\alpha} - A^{a}_{\alpha\rho\alpha^{'}}
-A^{a}_{\alpha\alpha^{'}\rho}-A^{a}_{\alpha\alpha^{'}\rho}
-A^{a}_{\alpha\rho\alpha^{'}})-\nn\\
-\partial_{\alpha}\partial_{\alpha^{'}} (A^{a}_{\alpha^{''}\rho\rho}
+A^{a}_{\rho\alpha^{''}\rho} + A^{a}_{\rho\rho\alpha^{''}})
-\partial_{\alpha}\partial_{\alpha^{''}} (A^{a}_{\alpha^{'}\rho\rho}
+A^{a}_{\rho\alpha^{'}\rho} + A^{a}_{\rho\rho\alpha^{'}})
+2 \partial_{\alpha^{'}}\partial_{\alpha^{''}} A^{a}_{\alpha\rho\rho}\}-\nn\\
    -{1 \over 8}\{\eta_{\alpha\alpha^{'}} [\partial^{2}( A^{a}_{\alpha^{''}\rho\rho}+
A^{a}_{\rho\alpha^{''}\rho}+A^{a}_{\rho\rho\alpha^{''}})-
\partial_{\alpha^{''}}\partial_{\rho} A^{a}_{\rho\lambda \lambda}-
\partial_{\lambda}\partial_{\rho} (A^{a}_{\rho \alpha^{''}\lambda}
+ A^{a}_{\rho \lambda\alpha^{''}})]+\nn\\
+\eta_{\alpha\alpha^{''}} [\partial^{2}( A^{a}_{\alpha^{'}\rho\rho}+
A^{a}_{\rho\alpha^{'}\rho}+A^{a}_{\rho\rho\alpha^{'}})-
\partial_{\alpha^{'}}\partial_{\rho} A^{a}_{\rho\lambda \lambda}-
\partial_{\lambda}\partial_{\rho} (A^{a}_{\rho \alpha^{'}\lambda}
+ A^{a}_{\rho \lambda\alpha^{'}})]+~~~\nn\\
+\eta_{\alpha^{'}\alpha^{''}}[ \partial^{2}( A^{a}_{\rho\alpha\rho}+
A^{a}_{\rho\rho\alpha})
-\partial_{\alpha} \partial_{\rho}( A^{a}_{\lambda\rho\lambda} +A^{a}_{\lambda\lambda} \rho)
-\partial_{\lambda} \partial_{\rho}( A^{a}_{\rho\alpha\lambda} + A^{a}_{\lambda\rho\alpha}
-2A^{a}_{\alpha\lambda\rho})] \}.
\eeqa
Summing these two pieces together we shall get the following free field equation of
motion for the rank-3 tensor gauge field:
\beqa
(H_{\alpha\alpha^{'}\alpha^{''}\gamma\gamma^{'}\gamma^{''}}
+c H^{~'}_{\alpha\alpha^{'}\alpha^{''}\gamma\gamma^{'}\gamma^{''}})
A^{a}_{\gamma\gamma^{'}\gamma^{''}}=
\partial^{2}( A^{a}_{\alpha \alpha^{'}\alpha^{''}}- {c\over 4}
A^{a}_{\alpha^{'}\alpha^{''}\alpha }
- {c\over 4} A^{a}_{\alpha^{''}\alpha \alpha^{'}})-\nn\\
-\partial_{\alpha} \partial_{\rho}(A^{a}_{\rho\alpha^{'}\alpha^{''}}
- {c\over 4}A^{a}_{\alpha^{'}\alpha^{''}\rho }
- {c\over 4} A^{a}_{\alpha^{''}\rho \alpha^{'}})
-{c\over 4}\partial_{\alpha^{'}} \partial_{\rho}(A^{a}_{\alpha\rho\alpha^{''}}
+A^{a}_{\alpha\alpha^{''}\rho} - A^{a}_{\rho\alpha\alpha^{''}})-\nn\\
-{c\over 4}\partial_{\alpha^{''}} \partial_{\rho}(A^{a}_{\alpha\rho^{'}\alpha}
+A^{a}_{\alpha\alpha^{'}\rho} - A^{a}_{\rho\alpha\alpha^{'}})
+{c\over 8}\partial_{\alpha}\partial_{\alpha^{'}} (A^{a}_{\alpha^{''}\rho\rho}
+A^{a}_{\rho\alpha^{''}\rho} + A^{a}_{\rho\rho\alpha^{''}})+\nn\\
+{c\over 8}\partial_{\alpha}\partial_{\alpha^{''}} (A^{a}_{\alpha^{'}\rho\rho}
+A^{a}_{\rho\alpha^{'}\rho} + A^{a}_{\rho\rho\alpha^{'}})
-{c\over 4}\partial_{\alpha^{'}}\partial_{\alpha^{''}} A^{a}_{\alpha\rho\rho}-\nn\\
-{c \over 8}\eta_{\alpha\alpha^{'}} (\partial^{2} A^{a}_{\alpha^{''}\rho\rho}-
\partial_{\alpha^{''}}\partial_{\rho} A^{a}_{\rho\lambda \lambda}
+2 \partial^{2}A^{a}_{\rho\rho\alpha^{''}}-2
\partial_{\lambda}\partial_{\rho} A^{a}_{\rho \lambda\alpha^{''}})-\nn\\
-{c \over 8}\eta_{\alpha\alpha^{''}} (\partial^{2} A^{a}_{\alpha^{'}\rho\rho}-
\partial_{\alpha^{'}}\partial_{\rho} A^{a}_{\rho\lambda \lambda}
+2 \partial^{2}A^{a}_{\rho\rho\alpha^{'}}-2
\partial_{\lambda}\partial_{\rho} A^{a}_{\rho \lambda\alpha^{'}})+\nn\\
+{1 \over 2}\eta_{\alpha^{'}\alpha^{''}}( \partial^{2} A^{a}_{\alpha\rho\rho}
-\partial_{\alpha} \partial_{\rho} A^{a}_{\rho\lambda\lambda}
-{c \over 2}\partial^{2} A^{a}_{\rho\rho\alpha}
+{c \over 2}\partial_{\alpha} \partial_{\rho} A^{a}_{\lambda\lambda\rho}
-{c \over 2}\partial_{\lambda} \partial_{\rho} A^{a}_{\alpha\lambda\rho}
+ {c \over 2}\partial_{\lambda} \partial_{\rho} A^{a}_{\lambda\rho\alpha})
=0, \nn
\eeqa
where $c=g^{'}_{3}/g_3$. Performing the gauge transformation (\ref{enhancedgaugetransformationrank3})
of the gauge field one can see that the terms which originate from
differential operators $\partial^{2},~$
$\partial_{\alpha} \partial_{\rho},~~$$\partial_{\alpha^{'}} \partial_{\rho}$
and $\partial_{\alpha^{''}} \partial_{\rho}$ in the above equation
cancel each other if we choose
$
g^{'}_{3} =  {4\over 3} g_{3}~.
$
The rest of the terms have the following form:
\beqa
(H_{\alpha\alpha^{'}\alpha^{''}\gamma\gamma^{'}\gamma^{''}}
+{4\over 3} H^{~'}_{\alpha\alpha^{'}\alpha^{''}\gamma\gamma^{'}\gamma^{''}})
\tilde{\delta} A^{a}_{\gamma\gamma^{'}\gamma^{''}}=\nn\\
+{1\over 3}\partial_{\alpha}\partial_{\alpha^{'}} \partial_{\rho}  \zeta^{a}_{\rho\alpha^{''}}
+{1\over 3}\partial_{\alpha}\partial_{\alpha^{''}} \partial_{\rho}  \zeta^{a}_{\rho\alpha^{'}}
-{4\over 3}\partial_{\alpha^{'}}\partial_{\alpha^{''}}\partial_{\rho} \zeta^{a}_{\rho\alpha}
+{2\over 3}\partial_{\alpha}\partial_{\alpha^{'}}\partial_{\alpha^{''}}  \zeta^{a}_{\rho\rho}\nn\\
-{1 \over 6}\eta_{\alpha\alpha^{'}} (2 \partial_{\rho} \partial^{2} \zeta^{a}_{\rho\alpha^{''}}
-4\partial_{\alpha^{''}}\partial_{\lambda} \partial_{\rho} \zeta^{a}_{\lambda \rho}
+2 \partial_{\alpha^{''}}\partial^{2} \zeta^{a}_{\rho\rho})\nn\\
-{1 \over 6}\eta_{\alpha\alpha^{''}} (2 \partial_{\rho} \partial^{2} \zeta^{a}_{\rho\alpha^{'}}
-4\partial_{\alpha^{'}}\partial_{\lambda} \partial_{\rho} \zeta^{a}_{\lambda \rho}
+2 \partial_{\alpha^{'}}\partial^{2} \zeta^{a}_{\rho\rho})\nn\\
+{1 \over 3}\eta_{\alpha^{'}\alpha^{''}}(\partial_{\rho} \partial^{2} \zeta^{a}_{\rho\alpha}
-\partial_{\alpha} \partial_{\lambda}\partial_{\rho} \zeta^{a}_{\lambda\rho})
\eeqa
and can be rewritten in the form which makes  the desired invariance explicit:
\beqa
(H_{\alpha\alpha^{'}\alpha^{''}\gamma\gamma^{'}\gamma^{''}}
+{4\over 3} H^{~'}_{\alpha\alpha^{'}\alpha^{''}\gamma\gamma^{'}\gamma^{''}})
\delta A^{a}_{\gamma\gamma^{'}\gamma^{''}}=
+{1\over 3}\partial_{\alpha}\partial_{\alpha^{'}}( \partial_{\rho}  \zeta^{a}_{\rho\alpha^{''}}
-\partial_{\alpha^{''}}  \zeta^{a}_{\rho\rho})\nn\\
+{1\over 3}\partial_{\alpha}\partial_{\alpha^{''}} (\partial_{\rho}  \zeta^{a}_{\rho\alpha^{'}}
-\partial_{\alpha^{'}}  \zeta^{a}_{\rho\rho})
-{4\over 3}\partial_{\alpha^{'}}\partial_{\alpha^{''}}(\partial_{\rho} \zeta^{a}_{\rho\alpha}
-\partial_{\alpha}  \zeta^{a}_{\rho\rho})\nn\\
-{1 \over 3}\eta_{\alpha\alpha^{'}} [ \partial^{2}( \partial_{\rho}  \zeta^{a}_{\rho\alpha^{''}}
-\partial_{\alpha^{''}}\zeta^{a}_{\rho\rho})+2
\partial_{\alpha^{''}}\partial_{\lambda} (\partial_{\lambda} \zeta^{a}_{ \rho\rho}-
\partial_{\rho}\zeta^{a}_{\rho\lambda})]\nn\\
-{1 \over 3}\eta_{\alpha\alpha^{''}} [ \partial^{2}( \partial_{\rho}  \zeta^{a}_{\rho\alpha^{'}}
-\partial_{\alpha^{'}}\zeta^{a}_{\rho\rho})+2
\partial_{\alpha^{'}}\partial_{\lambda} (\partial_{\lambda} \zeta^{a}_{ \rho\rho}-
\partial_{\rho}\zeta^{a}_{\rho\lambda})]\nn\\
+{1 \over 3}\eta_{\alpha^{'}\alpha^{''}} [ \partial^{2}(\partial_{ \rho}\zeta^{a}_{\rho\alpha}
-\partial_{\alpha}  \zeta^{a}_{\rho\rho})+
\partial_{\alpha}\partial_{\lambda} (\partial_{\lambda} \zeta^{a}_{ \rho\rho}-
\partial_{\rho}\zeta^{a}_{\rho\lambda})].
\eeqa
From that we see that if the gauge parameter satisfies the restriction (\ref{restrictionongaugeparameters})
the equation is indeed invariant with respect to a larger group of gauge
transformations
$
\tilde{\delta} A^{a}_{\mu \nu\lambda} =\partial_{\nu} \zeta^{a}_{\mu\lambda}+
\partial_{\lambda} \zeta^{a}_{\mu\nu},
$
because
\be
(H_{\alpha\alpha^{'}\alpha^{''}\gamma\gamma^{'}\gamma^{''}}
+{4\over 3} H^{~'}_{\alpha\alpha^{'}\alpha^{''}\gamma\gamma^{'}\gamma^{''}})
\tilde{\delta} A^{a}_{\gamma\gamma^{'}\gamma^{''}}
=\CH_{\alpha\alpha^{'}\alpha^{''} \gamma\gamma^{'}\gamma^{''}}(k)~
\tilde{\delta} A^{a}_{\gamma\gamma^{'}\gamma^{''}}=0.
\ee
The final form of the equation is
\beqa\label{freethirdrankequations}
\partial^{2}( A^{a}_{\alpha \alpha^{'}\alpha^{''}}
-{1\over 3}A^{a}_{\alpha^{'}\alpha^{''}\alpha }
- {1\over 3} A^{a}_{\alpha^{''}\alpha \alpha^{'}})
-\partial_{\alpha} \partial_{\rho}(A^{a}_{\rho\alpha^{'}\alpha^{''}}
- {1\over 3}A^{a}_{\alpha^{'}\alpha^{''}\rho }
- {1\over 3} A^{a}_{\alpha^{''}\rho \alpha^{'}})-\\
-{1\over 3}\partial_{\alpha^{'}} \partial_{\rho}(A^{a}_{\alpha\rho\alpha^{''}}
+A^{a}_{\alpha\alpha^{''}\rho} - A^{a}_{\rho\alpha\alpha^{''}})
-{1\over 3}\partial_{\alpha^{''}} \partial_{\rho}(A^{a}_{\alpha\rho\alpha^{'}}
+A^{a}_{\alpha\alpha^{'}\rho} - A^{a}_{\rho\alpha\alpha^{'}})+\nn\\
+{1\over 6}\partial_{\alpha}\partial_{\alpha^{'}} (A^{a}_{\alpha^{''}\rho\rho}
+A^{a}_{\rho\alpha^{''}\rho} + A^{a}_{\rho\rho\alpha^{''}})
+{1\over 6}\partial_{\alpha}\partial_{\alpha^{''}} (A^{a}_{\alpha^{'}\rho\rho}
+A^{a}_{\rho\alpha^{'}\rho} + A^{a}_{\rho\rho\alpha^{'}})
-{1\over 3}\partial_{\alpha^{'}}\partial_{\alpha^{''}} A^{a}_{\alpha\rho\rho}-\nn\\
-{1 \over 6}\eta_{\alpha\alpha^{'}} (\partial^{2} A^{a}_{\alpha^{''}\rho\rho}-
\partial_{\alpha^{''}}\partial_{\rho} A^{a}_{\rho\lambda \lambda}
+2 \partial^{2}A^{a}_{\rho\rho\alpha^{''}}-2
\partial_{\lambda}\partial_{\rho} A^{a}_{\rho \lambda\alpha^{''}})-\nn\\
-{1 \over 6}\eta_{\alpha\alpha^{''}} (\partial^{2} A^{a}_{\alpha^{'}\rho\rho}-
\partial_{\alpha^{'}}\partial_{\rho} A^{a}_{\rho\lambda \lambda}
+2 \partial^{2}A^{a}_{\rho\rho\alpha^{'}}-2
\partial_{\lambda}\partial_{\rho} A^{a}_{\rho \lambda\alpha^{'}})+\nn\\
+{1 \over 2}\eta_{\alpha^{'}\alpha^{''}}( \partial^{2} A^{a}_{\alpha\rho\rho}
-\partial_{\alpha} \partial_{\rho} A^{a}_{\rho\lambda\lambda}
-{2 \over 3}\partial^{2} A^{a}_{\rho\rho\alpha}
+{2 \over 3}\partial_{\alpha} \partial_{\rho} A^{a}_{\lambda\lambda\rho}
-{2 \over 3}\partial_{\lambda} \partial_{\rho} A^{a}_{\alpha\lambda\rho}
+ {2 \over 3}\partial_{\lambda} \partial_{\rho} A^{a}_{\lambda\rho\alpha})
=0\nn
\eeqa
and, it is invariant with respect to the group of gauge transformations
\be\label{fullgroupofextendedtransformation}
\delta A^{a}_{\mu\nu\lambda} =\partial_{\mu} \xi^{a}_{\nu\lambda},~~~~~~~~~
\tilde{\delta} A^{a}_{\mu \nu\lambda} =\partial_{\nu} \zeta^{a}_{\mu\lambda}+
\partial_{\lambda} \zeta^{a}_{\mu\nu},~~~~~~~~
\partial_{\rho}\zeta^{a}_{\rho\lambda}-\partial_{\lambda} \zeta^{a}_{ \rho\rho}=0.
\ee
The above invariance of the equation (\ref{freethirdrankequations}) with respect to the
transformations (\ref{fullgroupofextendedtransformation}) can be checked now directly without
referring to the previous analysis.

Let us now estimate, how many independent gauge parameters are at our disposal.
Because there are no restrictions on the symmetric gauge parameter $\xi^{a}_{\mu\nu}$,
we have ten independent gauge parameters in the four-dimensional space-time.
To estimate the amount of
independent gauge parameters in $\zeta^{a}_{ \mu\nu}$ one should solve
the restriction (\ref{restrictionongaugeparameters})
\beqa
\omega \zeta_{03} + \kappa \zeta_{33}+
\kappa (\zeta_{00}- \zeta_{11}-\zeta_{22}-\zeta_{33})=0,\nn\\
\omega \zeta_{01} + \kappa \zeta_{31}=0,\nn\\
\omega \zeta_{02} + \kappa  \zeta_{32} =0,\nn\\
\omega \zeta_{00}+ \kappa \zeta_{30} -
\omega (\zeta_{00}- \zeta_{11}-\zeta_{22}-\zeta_{33})=0,\nn
\eeqa
where $k^{\mu}=(\omega,0,0,\kappa)$, therefore
\beqa
\zeta_{00}=
\zeta_{11}+\zeta_{22} -{\omega \over \kappa } \zeta_{03} , ~~~~
\zeta_{31}= - {\omega \over \kappa }\zeta_{01},\nn\\
\zeta_{33}= - \zeta_{11}-\zeta_{22} -{\kappa \over \omega }\zeta_{03} ,~~~~
\zeta_{32} = - {\omega \over \kappa }\zeta_{02},\nn\\
\eeqa
and we have six independent gauge parameters\footnote{One can
also use a different set of independent parameters, in particular,
$
\zeta_{11},\zeta_{12}, \zeta_{13},\zeta_{22}, \zeta_{23}, \zeta_{33}.
$}
$
\zeta_{01},\zeta_{02}, \zeta_{03}, \zeta_{11}, \zeta_{22}, \zeta_{12}.
$
We shall present the free equation of motion (\ref{freethirdrankequations})
also in terms of field strength
tensors. The quadratic part of the Lagrangian is
\beqa\label{freeactionthreeprimesum}
{{\cal L}}_3 + {4 \over 3} {{\cal L}}^{'}_3 ~\vert_{quadratic}
=&-&{1\over 4}F^{a}_{\mu\nu,\lambda\rho}F^{a}_{\mu\nu,\lambda\rho}
-{1\over 8}F^{a}_{\mu\nu ,\lambda\lambda}F^{a}_{\mu\nu ,\rho\rho}\nn\\
&+&{1\over 3}F^{a}_{\mu\nu,\lambda\rho}F^{a}_{\mu\lambda,\nu\rho}+
{1\over 3} F^{a}_{\mu\nu,\nu\lambda}F^{a}_{\mu\rho,\rho\lambda}+
{1\over 3}F^{a}_{\mu\nu,\nu\lambda}F^{a}_{\mu\lambda,\rho\rho}=\nn\\
&=&{1 \over 2} A^{a}_{\alpha \alpha^{'}\alpha^{''}}
\CH_{\alpha \alpha^{'}\alpha^{''} \gamma\gamma^{'}\gamma^{''}}
A^{a}_{\gamma \gamma^{'} \gamma^{''} },
\eeqa
where
$
F^{a}_{\mu\nu ,\lambda \rho} =
\partial_{\mu} A^{a}_{\nu \lambda \rho} - \partial_{\nu} A^{a}_{\mu \lambda\rho}.
$
Its variation over the field
$A^{a}_{\nu \lambda \rho}$ gives
the free equation  (\ref{freethirdrankequations})  written in terms
of field strength tensor $F^{a}_{\mu\nu ,\lambda \rho}$:
\beqa\label{freeequation}
\partial_{\mu}F^{a}_{\mu\nu,\lambda\rho}
-{1\over 3}\partial_{\mu}F^{a}_{\mu\lambda,\nu\rho}
- {1\over 3} \partial_{\mu}F^{a}_{\mu\rho,\nu\lambda}
+ {1\over 3}\partial_{\mu}F^{a}_{\nu\lambda,\mu\rho}
+{1\over 3} \partial_{\mu}F^{a}_{\nu\rho,\mu\lambda} +\nn\\
+{1\over 3}\partial_{\lambda}F^{a}_{\nu\mu,\mu\rho}
+{1\over 3}\partial_{\rho}F^{a}_{\nu\mu,\mu\lambda}
+{1\over 6}\partial_{\lambda}F^{a}_{\nu\rho,\mu\mu}
+{1\over 6}\partial_{\rho}F^{a}_{\nu\lambda,\mu\mu}-\nn\\
-\eta_{\lambda\nu} ({1\over 3}\partial_{\mu}F^{a}_{\mu\sigma,\sigma\rho}
+{1\over 6}\partial_{\mu}F^{a}_{\mu\rho,\sigma\sigma} )
-\eta_{\nu\rho} ({1\over 3}\partial_{\mu}F^{a}_{\mu\sigma,\sigma\lambda}
+{1\over 6}\partial_{\mu}F^{a}_{\mu\lambda,\sigma\sigma} )+\nn\\
+ \eta_{\lambda\rho}({1\over 2} \partial_{\mu}F^{a}_{\mu\nu,\sigma\sigma}
- {1\over 3}\partial_{\mu}F^{a}_{\mu\sigma,\sigma\nu}
+{1\over 3}\partial_{\mu}F^{a}_{\nu\sigma,\sigma\mu})=
0.
\eeqa
In summary, we have the Lagrangian (\ref{thirdranktensorlagrangian}) for the third-rank gauge field
$A^{a}_{\mu\nu\lambda}$ and the corresponding free field equation of motion (\ref{freethirdrankequations})/
(\ref{freeequation})  invariant with respect to the
gauge transformations (\ref{fullgroupofextendedtransformation}) .

\section{\it Propagating Modes of Rank-3 Gauge Field}

The aim of this section is to analyze the free field equation
(\ref{freethirdrankequations})/
(\ref{freeequation}) for the rank-3 gauge field. It is convenient to decompose the rank-3
gauge field into irreducible pieces. The gauge field
$A^{a}_{\gamma\gamma^{'}\gamma^{''}}$ is
symmetric over the last two indices $\gamma^{'} \leftrightarrow \gamma^{''}$ and has
no symmetries with respect to the index $\gamma$. Let us consider the transformation $T$
of the form \cite{Barrett:2007nn}
\beqa\label{dualityfieldtransformations}
  \begin{array}{ll}
A^{T}_{\mu\lambda_1} =  A_{\lambda_1\mu}   ,  \\
A^{T}_{\mu\lambda_1\lambda_2} =
{2\over 3}(A_{\lambda_1\mu\lambda_2} + A_{\lambda_2\mu\lambda_1})
-{1\over 3} A_{\mu\lambda_1\lambda_2}, \\
.......................................
\end{array}
\eeqa
It has the property of the standard transposition
$
(A^{T})^T = A
$
and allows to define symmetric $A^S$ and anti-symmetric $A^A$ tensors as
$$
A^{S}={1\over 2}(A + A^{T}),~~~~A^{A}={1\over 2}(A - A^{T}).
$$
In the case of the rank-3 gauge field they are
\be\label{symmtrantisymm}
A^{S}_{\mu\lambda_1\lambda_2} =
{1\over 3}(A_{\mu\lambda_1\lambda_2}+A_{\lambda_1\mu\lambda_2} + A_{\lambda_2\mu\lambda_1}),~~~
A^{A}_{\mu\lambda_1\lambda_2} = {2\over 3} A_{\mu\lambda_1\lambda_2}
-{1\over 3}(A_{\lambda_1\mu\lambda_2} + A_{\lambda_2\mu\lambda_1}).
\ee
One should also define vector fields associated with rank-3 tensor field:
\be\label{traces}
B_{\mu} = A_{\mu \lambda\lambda}, ~~~C_{\mu} = A_{\lambda\lambda\mu },~~~
D_{\mu} = \partial_{\lambda} \partial_{\rho}
A_{\mu\lambda\rho},~~~E_{\mu} = \partial_{\lambda} \partial_{\rho}
A_{\lambda\rho\mu}.
\ee
Equation for these fields follow from our main equation (\ref{freethirdrankequations})/
(\ref{freeequation}), if one takes its trace
\beqa
 (\eta_{\mu\nu}\partial^{2} - \partial_{\mu} \partial_{\nu}) ({7\over 8}B_{\nu}-C_{\nu})=
(D_{\mu}- E_{\mu}).
\eeqa
These equations show that these vector fields (\ref{traces})
fulfill Maxwell equation.
Our aim  is to find explicit solutions for all these fields. This will allow
to clarify the physical content of the equation (\ref{freethirdrankequations})/
(\ref{freeequation}) and the propagating modes which it describes.

A convenient way to solve the free equation of motion is to consider it in momentum representation
\be\label{basicequationrank3}
(H_{\alpha\alpha^{'}\alpha^{''}\gamma\gamma^{'}\gamma^{''}}
+{4\over 3} H^{'}_{\alpha\alpha^{'}\alpha^{''}\gamma\gamma^{'}\gamma^{''}})
A^{a}_{\gamma\gamma^{'}\gamma^{''}}
=
\CH_{\alpha\alpha^{'}\alpha^{''} \gamma\gamma^{'}\gamma^{''}}(k)~
A^{a}_{\gamma\gamma^{'}\gamma^{''}} =0,
\ee
as we did in the case of the rank-2 gauge field in section three.
The matrix operator $\CH_{\alpha\alpha^{'}\alpha^{''}\gamma\gamma^{'}\gamma^{''}}(k)$
is the sum of $H$ and $H^{'}$  given by (\ref{rank3form1}) and (\ref{rank3form2}) and
in the four-dimensional space-time it is a square matrix  $40 \times 40$. Indeed, the gauge field
$A^{a}_{\gamma\gamma^{'}\gamma^{''}}$ is
symmetric over the last two indices $\gamma^{'} \leftrightarrow \gamma^{''}$ and has
no symmetries with respect to the index $\gamma$, thus the multi-index
$N \equiv (\gamma,\gamma^{'},\gamma^{''}) $ runs $4 \times 10=40$ values and
the matrix $\CH_{NM}$ is  $40 \times 40$.
It is convenient to represent the gauge field in the form of four symmetric matrices
$A_{\gamma\gamma^{'}\gamma^{''}}=(A_{0\gamma^{'}\gamma^{''}},...,A_{3\gamma^{'}\gamma^{''}})=
(e_{0\gamma^{'}\gamma^{''}},...,e_{3\gamma^{'}\gamma^{''}}) exp\{i k x\}$
\beqa
e_{\gamma\gamma^{'}\gamma^{''}}=
\left(\left(\begin{array}{cccc}
e_{000}&e_{001}&e_{002}&e_{003} \\
e_{010}&e_{011}&e_{012}&e_{013} \\
e_{020}&e_{021}&e_{022}&e_{023} \\
e_{030}&e_{031}&e_{032}&e_{033} \\
\end{array} \right),...,
\left(\begin{array}{cccc}
e_{300}&e_{301}&e_{302}&e_{303} \\
e_{310}&e_{311}&e_{312}&e_{313} \\
e_{320}&e_{321}&e_{322}&e_{323} \\
e_{330}&e_{331}&e_{332}&e_{333} \\~
\end{array} \right)\right)
\eeqa
each of which has ten independent components.
In the reference frame, where
$k^{\gamma}=(\omega,0,0,k)$, the matrix $\CH_{NM}$ has a particularly simple form.
If $\omega^2 - k^2 \neq 0$, the rank of the 40-dimensional
matrix
$
\CH_{NM}(k)
$
is equal to $rank ~\CH\vert_{\omega^2 - k^2 \ne 0}=25$
and the number of linearly independent solutions is $40-25=15$.
These are pure gauge fields (\ref{fullgroupofextendedtransformation})
\be\label{puregaugepotentialsrank3}
e_{\gamma\gamma^{'}\gamma^{''}}=
k_{\gamma}~\xi_{\gamma^{'}\gamma^{''}}+
k_{\gamma^{'} }\zeta_{\gamma\gamma^{''}} +k_{\gamma^{''} }\zeta_{\gamma\gamma^{'}}
\ee
with ten  independent
gauge parameters $\xi_{\gamma^{'}\gamma^{''}}$ and five  independent
gauge parameters $\zeta_{\gamma\gamma^{'}}.$

When $\omega^2 - k^2 = 0$, then the rank of the matrix
$
\CH_{NM}(k)
$
drops and is equal to $rank \CH\vert_{\omega^2 - k^2 = 0}  =18$.
This leaves us with $40-18=22$ solutions. These are 15+1=16 solutions,
the pure gauge fields (\ref{fullgroupofextendedtransformation}), (\ref{puregaugepotentialsrank3})
and six solutions representing propagating modes. On the mass-shell the number
of pure gauge fields increases by one unit: instead
of the pure gauge field
\beqa\label{decouplingpuregaugefields}
&e_{\gamma\gamma^{'}\gamma^{''}} =
k_{ \gamma^{'} } (e^{(1)}_{\gamma} e^{(2)}_{\gamma^{''}} +
                            e^{(2)}_{\gamma} e^{(1)}_{\gamma^{''}})
+k_{ \gamma^{''} }(e^{(1)}_{\gamma} e^{(2)}_{\gamma^{'}} +
                        e^{(2)}_{\gamma} e^{(1)}_{\gamma^{'}})
\eeqa
two new linearly independent solutions appear
\be
\rightarrow
\begin{array}{ll}
e_{\gamma\gamma^{'}\gamma^{''}}^{ ' }=
k_{ \gamma^{'} } e^{(1)}_{\gamma} e^{(2)}_{\gamma^{''}}
    +k_{ \gamma^{''} }e^{(1)}_{\gamma} e^{(2)}_{\gamma^{'}} \\
e_{\gamma\gamma^{'}\gamma^{''}}^{ '' }=
 k_{ \gamma^{'} } e^{(2)}_{\gamma} e^{(1)}_{\gamma^{''}}
+k_{ \gamma^{''} }e^{(2)}_{\gamma} e^{(1)}_{\gamma^{'}}  \\
\end{array}\nn
\ee
where
$
e^{(1)}_{\mu}=(0,1,0,0),~~  e^{(2)}_{\mu}= (0,0,1,0).
$
Therefore on the mass-shell we have sixteen pure gauge fields and
six propagating modes 22-16=6.
The first two solutions are:
\beqa\label{physicalmodesrank3pm3}
e_{\gamma\gamma^{'}\gamma^{''}}^{(1)}&=&\left(
0,
\left(\begin{array}{cccc}
0&0&0&0 \\
0&1&0&0 \\
0&0&-1&0 \\
0&0&0&0 \\
\end{array} \right),
\left(\begin{array}{cccc}
0&0&0&0 \\
0&0&-1&0 \\
0&-1&0&0 \\
0&0&0&0 \\
\end{array} \right),0\right)
\nn
\\
e_{\gamma\gamma^{'}\gamma^{''}}^{(2)}&=&\left(
0,
\left(\begin{array}{cccc}
0&0&0&0 \\
0&0&1&0 \\
0&1&0&0 \\
0&0&0&0 \\
\end{array} \right),
\left(\begin{array}{cccc}
0&0&0&0 \\
0&1&0&0 \\
0&0&-1&0 \\
0&0&0&0 \\
\end{array} \right),0\right).\nn
\eeqa
These are traceless tensors $(B_{\mu}= C_{\mu}=D_{\mu}=E_{\mu}=0)$.
Their linear combinations describe positive norm states with
helicities $\lambda=\pm 3$, because one can represent these solutions
as a direct product of
helicity-one and helicity-two tensors
$
e_{\gamma\gamma^{'}\gamma^{''}}=e^{\pm 1}_{\gamma} \otimes e^{\pm 2}_{\gamma^{'}\gamma^{''}}.
$
The next two solutions are:
\beqa\label{physicalmodesrank3pm1second}
&e_{\gamma\gamma^{'}\gamma^{''}}^{(5)}=
\left(
0,
\left(\begin{array}{cccc}
0&0&0&0 \\
0&1&0&0 \\
0&0&-1&0 \\
0&0&0&0 \\
\end{array} \right),
0,
0\right) -{1 \over 3}(
 \eta_{\gamma \gamma^{'} } e^{(1)}_{\gamma^{''}}
+\eta_{\gamma \gamma^{''} }e^{(1)}_{\gamma^{'}})
\\
&e_{\gamma\gamma^{'}\gamma^{''}}^{(6)}=\left(
0,0,
\left(\begin{array}{cccc}
0&0&0&0 \\
0&-1&0&0 \\
0&0&1&0 \\
0&0&0&0 \\
\end{array} \right),
0\right) -{1 \over 3}
(\eta_{\gamma \gamma^{'} } e^{(2)}_{\gamma^{''}}
+\eta_{\gamma \gamma^{''} }e^{(2)}_{\gamma^{'}}).\nn
\eeqa
In accordance with (\ref{physicalmodesrank3pm1second}) we have
$$
B^{(5,6)}_{\mu} = -{2 \over 3}e^{(1,2)}_{\mu},~~~C^{(5,6)}_\mu =-{2 \over 3}e^{(1,2)}_{\mu},~~~
D_{\mu}=E_{\mu}=0
$$
and they fulfill the free Maxwell equation (\ref{freeequation}).
Their linear combinations describe positive norm
states of helicities $\lambda =\pm 1$.
The last two solutions are:
\beqa\label{physicalmodesrank3pm1}
e_{\gamma\gamma^{'}\gamma^{''}}^{(3)}=\left(
0,
\left(\begin{array}{cccc}
0&0&0&0 \\
0&0&0&0 \\
0&0&1&0 \\
0&0&0&0 \\
\end{array} \right),
0,
0\right) - {1\over 8}e^{(1)}_{\gamma} \eta_{\gamma^{'} \gamma^{''}}
\nn\\
e_{\gamma\gamma^{'}\gamma^{''}}^{(4)}=\left(
0,
0,
\left(\begin{array}{cccc}
0&0&0&0 \\
0&1&0&0 \\
0&0&0&0 \\
0&0&0&0 \\
\end{array} \right),0\right)- {1\over 8}
e^{(2)}_{\gamma}\eta_{\gamma^{'} \gamma^{''}}.
\eeqa
In accordance with solutions (\ref{physicalmodesrank3pm1}) we have
$$
B^{(3,4)}_{\mu} = {1 \over 2}e^{(1,2)}_{\mu},~~~C^{(3,4)}_\mu =-{1 \over 8}e^{(1,2)}_{\mu},~~~
D_{\mu}=E_{\mu}=0
$$
and they also fulfill the free Maxwell equation (\ref{freeequation}).
Their linear combinations describe positive norm states of helicities
$\lambda =\pm 1$. As one can check, the last solutions can not be decomposed
into symmetric and antisymmetric pieces (\ref{symmtrantisymm}).
The reason is that the kinetic
operator $\CH_{\alpha\alpha^{'}\alpha^{''}\gamma\gamma^{'}\gamma^{''}}$ in (\ref{basicequationrank3})
can not be represented as a sum of symmetric and anti-symmetric
operators. It has non-diagonal matrix elements and these solutions are
a mixture of the both symmetries. This is a new phenomenon which appears in the case of rank3
gauge field. In the case of rank-2 gauge field the kinetic operator
$\CH_{\alpha\alpha^{'}\gamma\gamma^{'}}$ in (\ref{quadraticform}) can be
decomposed into symmetric and anti-symmetric pieces, as we have seen in section
three.

Thus the general solution of the equation on the mass-shell is:
\be\label{gensolutionrank3}
e_{\gamma\gamma^{'}\gamma^{''}}=
k_{\gamma}~\xi_{\gamma^{'}\gamma^{''}}+
k_{\gamma^{'} }\zeta_{\gamma\gamma^{''}} +k_{\gamma^{''} }\zeta_{\gamma\gamma^{'}}+
\sum^{6}_{i=1}c_{i}e^{(i)}_{\gamma\gamma^{'}\gamma^{''}},
\ee
where $c_{i}$ are arbitrary constants.
Thus we see that there are six propagating modes
of {\it helicity-three and a doublet of helicity-one charged gauge bosons:
$\lambda = \pm 3, \pm 1, \pm1$}.

It is also interesting to see what happens if we consider  free field
equation (\ref{freethirdrankequations})/
(\ref{freeequation}) in $\CD$-dimensional space-time. As one can see the number
of potentially negative norm states increases as $(3\CD^3 -5 \CD +4)/ 2$
while the number of gauge parameters grows as $\CD^2$. Only in 3+1 dimensional
space-time there is a chance for full cancelation of negative norm states, and,
indeed, as we have seen, the particle spectrum is physical in 3+1 dimensions.
In five dimensions  the matrix $\CH_{NM}$ has dimension $75 \times 75$.
In the reference frame, where $k^{\gamma}=(\omega,0,0,k)$ and
$\omega^2 - k^2 \neq 0$, the rank  of the
matrix
$
\CH_{NM}(k)
$
is equal to $rank\CH\vert_{\omega^2 - k^2 \ne 0}=50$
and the number of linearly independent solutions is $25$.
These are pure gauge fields (\ref{fullgroupofextendedtransformation}),
(\ref{puregaugepotentialsrank3}) with fifteen  independent
gauge parameters $\xi_{\gamma^{'}\gamma^{''}}$ and ten independent
gauge parameters $\zeta_{\gamma\gamma^{'}}.$ When $\omega^2 - k^2 = 0$,
then $rank \CH\vert_{\omega^2 - k^2 = 0}  =30$.
This leaves us with $45$ solutions. These are 25 pure gauge
solutions and 20 new solutions representing propagating modes.
Only 18 modes can be positive definite.

Let us also consider equations for the higher-rank tensor gauge fields.
The Lagrangian
form for the rank-4 gauge field is a sum of the following two terms
(s=3 in (\ref{fulllagrangian1}), (\ref{secondfulllagrangian})) \cite{Savvidy:2005vm}:
\beqa
\CL_4
=&-&
{1 \over 4}
G_{\mu\nu, \rho\sigma\lambda}G_{\mu\nu, \rho\sigma\lambda}
-\frac{3}{8}
G_{\mu\nu, \sigma\rho\rho} G_{\mu\nu, \sigma\lambda\lambda}
-\frac{3}{4}
G_{\mu\nu, \rho\sigma}G_{\mu\nu, \rho\sigma\lambda\lambda}
\nonumber
\\
&
-&\frac{3}{16}
G_{\mu\nu, \rho\rho}G_{\mu\nu, \sigma\sigma\lambda\lambda}
-\frac{3}{8}
G_{\mu\nu, \rho}G_{\mu\nu, \rho\sigma\sigma\lambda\lambda}
-\frac{1}{16}
G_{\mu\nu}G_{\mu\nu, \rho\rho\sigma\sigma\lambda\lambda}
\eeqa
and
\beqa
\CL_4'
&=&
\frac{1}{4}
G_{\mu\nu, \rho\sigma\lambda}G_{\mu\rho, \nu\sigma\lambda}
+
\frac{1}{4}G_{\mu\nu, \nu\rho\sigma}G_{\mu\lambda, \lambda\rho\sigma}
+\frac{1}{8}
G_{\mu\nu, \rho\sigma\sigma}G_{\mu\rho, \nu\lambda\lambda}
+ \frac{1}{2}
G_{\mu\nu, \rho\sigma\sigma}G_{\mu\lambda, \nu\rho\lambda}
\nonumber
\\
&+&\frac{1}{8}
G_{\mu\nu, \nu\rho\rho}G_{\mu\sigma, \sigma\lambda\lambda}
+ \frac{1}{2}
G_{\mu\nu, \rho\sigma}G_{\mu\rho, \nu\sigma\lambda\lambda}
+ \frac{1}{2}
G_{\mu\nu, \rho\sigma}G_{\mu\lambda, \nu\rho\sigma\lambda}
+\frac{1}{4}
G_{\mu\nu, \rho\rho}G_{\mu\sigma, \nu\sigma\lambda\lambda}
\nonumber
\\
&+&\frac{1}{8}
G_{\mu\nu, \nu\rho}G_{\mu\rho, \lambda\lambda\sigma\sigma}
+ \frac{1}{2}
G_{\mu\nu, \nu\rho}G_{\mu\lambda, \rho\lambda\sigma\sigma}
+\frac{1}{8}
G_{\mu\nu, \rho}G_{\mu\rho, \nu\lambda\lambda\sigma\sigma}
+\frac{1}{2}
G_{\mu\nu, \rho}G_{\mu\lambda, \nu\rho\lambda\sigma\sigma}
\nonumber
\\
&
+&\frac{1}{8}
G_{\mu\nu, \nu}G_{\mu\rho, \rho\sigma\sigma\lambda\lambda}
+\frac{1}{8}
G_{\mu\nu}G_{\mu\rho, \nu\rho\sigma\sigma\lambda\lambda}.
\eeqa
Deriving field equations from this Lagrangian one can see that the free equation
of motion has two solutions which describe the propagating $\lambda =\pm 4$
helicity states only if
\be
g^{'}_4 = {3 \over 2} g_4~.
\ee
In the case of rank-(s+1) gauge field, as it follows  from
(\ref{fulllagrangian1}) and (\ref{secondfulllagrangian}),  the free equation
of motion has two solutions  describing the propagating $\lambda =\pm (s+1)$
helicity states only if
\be
g^{'}_{s+1} = {2s \over s+1} g_{s+1},
\ee
where $s=0,1,2,...$ and  $g_1=g_{YM} $. Therefore the Lagrangian
(\ref{generalgaugedensity}) has
the following form:
\be\label{fulllagrangian3}
{{\cal L}} = {{\cal L}}_{YM} + g_{2}({{\cal L}}_{2}+ {{\cal L}}^{'}_{2})
+g_{3}({{\cal L}}_{3}+ {4\over 3}{{\cal L}}^{'}_{3})+...+
g_{s+1}({{\cal L}}_{s+1}+ {2s\over s+1}{{\cal L}}^{'}_{s+1})+...,
\ee
where the coupling constants $g_{s+1}$ still remain undefined.
Therefore, let us consider the dependence of the Lagrangian on the coupling
constant $g_2$.  As we shall
see the coupling constant $g_2$ can be eliminated from the
Lagrangian by redefinition of fields and other coupling constants. Indeed,
let us define the transformation of tensor gauge fields as follows:
\be
A^{a}_{\mu\lambda_1 ... \lambda_{s}} \rightarrow  {1 \over g^{s/2}_{2}}~
A^{a}_{\mu\lambda_1 ... \lambda_{s}}.
\ee
This transformation should be complemented by the transformation of gauge
parameters
\be
\xi^{a}_{\lambda_1 ... \lambda_{s}} \rightarrow  {1 \over g^{s/2}_{2}}~
\xi^{a}_{\lambda_1 ... \lambda_{s}}
\ee
in order to protect the extended gauge transformations (\ref{polygauge}).
In that case the extended field strength tensors will also transform
homogeneously:
\be
G^{a}_{\mu\nu,\lambda_1 ... \lambda_{s}} \rightarrow  {1 \over g^{s/2}_{2}}~
G^{a}_{\mu\nu,\lambda_1 ... \lambda_{s}}
\ee
and the invariant forms will transform as follows:
\be
\CL_s  \rightarrow  {1 \over g^{s-1}_{2}}~
\CL_s .
\ee
Therefore the Lagrangian will take the form
\beqa\label{fulllagrangian4}
{{\cal L}} &=& {{\cal L}}_{YM} +   {{\cal L}}_{2}+ {{\cal L}}^{'}_{2}
+{g_{3}\over g^{2}_{2}}  ({{\cal L}}_{3}+ {4\over 3}{{\cal L}}^{'}_{3})
+{g_{4}\over g^{3}_{2}}  ({{\cal L}}_{4}+ {3\over 2}{{\cal L}}^{'}_{4})+...
\rightarrow\nn\\
&\rightarrow&{{\cal L}}_{YM} +   {{\cal L}}_{2}+ {{\cal L}}^{'}_{2}
+ g_{3} ({{\cal L}}_{3}+ {4\over 3}{{\cal L}}^{'}_{3})
+ g_{4} ({{\cal L}}_{4}+ {3\over 2}{{\cal L}}^{'}_{4})+
...
\eeqa
and the coupling constant $g_2$ is fully eliminated from the theory.
This cannot be done with the coupling constant $g_3$.

Summarizing our findings we can state that the Lagrangian $\CL$ describes the interacting
system of gauge bosons of increasing helicities. The system has
Yang-Mills gauge boson on the first level (s=0), the helicity-two and -zero gauge bosons
on the second level (s=1) and the helicity-three and a doublet of helicity-one
gauge bosons on the third level
(s=2).

The particle spectrum on higher levels is not yet known completely and
to find it out remains a challenging
problem. The problem consists in finding out the
value of the coupling constant  $g^{'}_{s+1}$ at which
the corresponding free field equation for the rank-(s+1) gauge field is free from
propagating negative norm states. As we have found for
$$
g^{'}_{s+1} = {2s \over s+1} g_{s+1},~~~~~~s=0,1,2,...
$$
($g_1=g_{YM})$ there are two solutions which describe the propagating
positive norm states of helicities $\lambda =\pm (s+1)$.
But the difficulty in finding out {\it all} propagating modes for this value  of the
coupling constant $g^{'}_{s+1}$ lies in the fact that the number of
field components dramatically
increases with the rank of the tensor gauge field: in the case of rank-2 gauge field
we had sixteen components and in the case considered in this article for
the rank-3 gauge field we had to analyze an equation with forty components.
The presented analysis shows that, most probably, the full system is
unitary for all higher-rank non-Abelian tensor gauge fields.

In conclusion let us discuss the relation between the present field
theoretical model
and the Coleman-Mandula theorem which imposes strict restrictions on
the possible
theories consistent with the fundamental principles of quantum field
theory \cite{Coleman:1967ad}.
The results of the  Coleman-Mandula paper were generally accepted as most
powerful in a series of "no-go" theorems, destroying the hope for a
fusion between
internal symmetries and the Poincar\'e group. It is applicable  if
five conditions formulated in the Coleman-Mandula article are  hold.
One of these conditions - {\it Particle-finiteness condition} - states that:
"(2) For any finite $M > 0$, there is only a finite
number of one-particle states with mass less than M."  The equivalent
formulation can also be found in the book of Wess and Bagger \cite{Wess:1992cp}
and in an important discussion in the article \cite{Gross:1988ue}.

The particle-finiteness condition is not
applicable to the field-theoretical model studied in the present article because
there are $\alpha )$  massless particles in the spectrum and $\beta )$
the number of massless particles is  infinite, in which case {\it the
theorem does not apply}.

There are well known cases when the Coleman-Mandula theorem is not applicable.
First of all it is the case already mentioned in Coleman-Mandula article,
the so called "infinite-supermultiplet theories"  and the second case is
the supersymmetric extension of the Poincar\'e algebra
\cite{Haag:1974qh,Wess:1992cp}.
Our model belongs to the first exceptional case.

In this article we  study the spectrum of the non-Abelian
tensor gauge fields and describe in details
the helicity content of these tensor fields. These studies comprise a necessary
step in any serious investigation, without which it is impossible to
{\it accept or reject
any theory}. The article does not contain claims that the suggested
model is a fully consistent
field theoretical model of interacting non-Abelian tensor gauge fields, but
takes the necessary steps in order to get an answer to the above question.

I would like to thank  Ignatios  Antoniadis,
Ludwig Faddeev  and Emmanuel Floratos for stimulating
discussions and CERN Theory Division, where part of this work was completed,
for hospitality.
The work was supported by ENRAGE (European Network on Random
Geometry), Marie Curie Research Training Network, contract MRTN-CT-2004-
005616.

\section{\it Appendix A}

The quadratic form $H^{~'}_{\alpha\alpha^{'}\alpha^{''}\gamma\gamma^{'}\gamma^{''}}$
can be extracted from (\ref{freeactionthreeprime})
and should be symmetrized  over
$\alpha^{'} \leftrightarrow\alpha^{''}$,
$\gamma^{'} \leftrightarrow \gamma^{''}$ and over the exchange of two sets
of indices
$\alpha \alpha^{'}\alpha^{''} \leftrightarrow \gamma\gamma^{'}\gamma^{''}$
so that in the momentum representation it has the form
\beqa
H^{~'}_{\alpha\alpha^{'}\alpha^{''}\gamma\gamma^{'}\gamma^{''}}(k)=
{ k^2 \over 8}\{ &+&\eta_{\alpha\alpha^{'}}
(\eta_{\alpha^{''}\gamma} \eta_{\gamma^{'} \gamma^{''}}
+\eta_{\alpha^{''}\gamma^{'}} \eta_{\gamma \gamma^{''}}
+\eta_{\alpha^{''}\gamma^{''}} \eta_{\gamma\gamma^{'} })\nn\\
&+&\eta_{\alpha\alpha^{''}}
(\eta_{\alpha^{'}\gamma} \eta_{\gamma^{'} \gamma^{''}}
+\eta_{\alpha^{'}\gamma^{'}} \eta_{\gamma \gamma^{''}}
+\eta_{\alpha^{'}\gamma^{''}} \eta_{\gamma\gamma^{'} })\nn\\
&+&\eta_{\alpha\gamma^{'}}
(\eta_{\alpha^{'}\gamma} \eta_{\alpha^{''} \gamma^{''}}
+\eta_{\alpha^{'}\gamma^{''}} \eta_{\alpha^{''}\gamma }
+\eta_{\alpha^{'}\alpha^{''}} \eta_{\gamma\gamma^{''} })\nn\\
&+&\eta_{\alpha\gamma^{''}}
(\eta_{\alpha^{'}\gamma} \eta_{\alpha^{''} \gamma^{'}}
+\eta_{\alpha^{'}\gamma^{'}} \eta_{\alpha^{''}\gamma }+
\eta_{\alpha^{'}\alpha^{''}} \eta_{\gamma \gamma^{'}})~\}\nn\\
             -{ 1\over 8}\{&+&k_{\alpha}k_{\alpha^{'}}
(\eta_{\alpha^{''}\gamma} \eta_{\gamma^{'} \gamma^{''}}
+\eta_{\alpha^{''}\gamma^{'}} \eta_{\gamma\gamma^{''} }
+\eta_{\alpha^{''}\gamma^{''}} \eta_{\gamma\gamma^{'} })\nn\\
&+&k_{\alpha}k_{\alpha^{''}}
(\eta_{\alpha^{'}\gamma} \eta_{\gamma^{'}\gamma^{''} }
+\eta_{\alpha^{'}\gamma^{'}} \eta_{\gamma \gamma^{''}}
+\eta_{\alpha^{'}\gamma^{''}} \eta_{\gamma \gamma^{'}})\nn\\
             &+&k_{\alpha}k_{\gamma^{'}}
(\eta_{\alpha^{'}\gamma} \eta_{\alpha^{''} \gamma^{''}}
+\eta_{\alpha^{'}\gamma^{''}} \eta_{\alpha^{''}\gamma }
+\eta_{\alpha^{'}\alpha^{''}} \eta_{\gamma\gamma^{''} })\nn\\
&+&k_{\alpha}k_{\gamma^{''}}
(\eta_{\alpha^{'}\gamma} \eta_{\alpha^{''} \gamma^{'}}
+\eta_{\alpha^{'}\gamma^{'}} \eta_{\alpha^{''}\gamma }
+\eta_{\alpha^{'}\alpha^{''}} \eta_{\gamma \gamma^{'}})\nn\\
&+&k_{\gamma}k_{\alpha^{'}}
(\eta_{\alpha\gamma^{'}} \eta_{\alpha^{''} \gamma^{''}}
+\eta_{\alpha\gamma^{''}} \eta_{\alpha^{''} \gamma^{'}}
+\eta_{\alpha\alpha^{''}} \eta_{\gamma^{'}\gamma^{''} })\nn\\
&+&k_{\gamma}k_{\alpha^{''}}
(\eta_{\alpha\gamma^{'}} \eta_{\alpha^{'} \gamma^{''}}
+\eta_{\alpha\gamma^{''}} \eta_{\alpha^{'}\gamma^{'} }
+\eta_{\alpha\alpha^{'}} \eta_{\gamma^{'}\gamma^{''} })\nn\\
&+&k_{\gamma}k_{\gamma^{'}}
(\eta_{\alpha\alpha^{'}} \eta_{\alpha^{''} \gamma^{''}}
+\eta_{\alpha\alpha^{''}} \eta_{\alpha^{'} \gamma^{''}}
+\eta_{\alpha\gamma^{''}} \eta_{\alpha^{'}\alpha^{''} })\nn\\
&+&k_{\gamma}k_{\gamma^{''}}
(\eta_{\alpha\alpha^{'}} \eta_{\alpha^{''} \gamma^{'}}
+\eta_{\alpha\alpha^{''}} \eta_{\alpha^{'}\gamma^{'} }
+\eta_{\alpha\gamma^{'}} \eta_{\alpha^{'}\alpha^{''} })~\}\nn\\
   +{ 1\over 4}\{&+&\eta_{\alpha\gamma} (k_{\alpha^{'}}k_{\gamma^{'}}
\eta_{\alpha^{''} \gamma^{''}} + k_{\alpha^{'}}k_{\gamma^{''}}
\eta_{\alpha^{''} \gamma^{'}} + k_{\alpha^{''}}k_{\gamma^{'}}
\eta_{\alpha^{'} \gamma^{''}} \nn\\
&+&k_{\alpha^{''}}k_{\gamma^{''}}
\eta_{\alpha^{'} \gamma^{'}} + k_{\alpha^{'}}k_{\alpha^{''}}
\eta_{ \gamma^{'}\gamma^{''}} +k_{\gamma^{'}}k_{\gamma^{''}}
\eta_{\alpha^{'}\alpha^{''} })~\}.
\eeqa
or combining some of the terms together we shall get an equivalent form
\beqa
H^{~'}_{\alpha\alpha^{'}\alpha^{''}\gamma\gamma^{'}\gamma^{''}}(k)=
{1\over 8}\{ &+& (k^2 \eta_{\alpha\alpha^{'}}-k_{\alpha}k_{\alpha^{'}})
(\eta_{\alpha^{''}\gamma} \eta_{\gamma^{'} \gamma^{''}}
+\eta_{\alpha^{''}\gamma^{'}} \eta_{\gamma \gamma^{''}}
+\eta_{\alpha^{''}\gamma^{''}} \eta_{\gamma\gamma^{'} })\nn\\
&+& (k^2 \eta_{\alpha\alpha^{''}} -k_{\alpha}k_{\alpha^{''}})
(\eta_{\alpha^{'}\gamma} \eta_{\gamma^{'} \gamma^{''}}
+\eta_{\alpha^{'}\gamma^{'}} \eta_{\gamma \gamma^{''}}
+\eta_{\alpha^{'}\gamma^{''}} \eta_{\gamma\gamma^{'} })\nn\\
&+&(k^2 \eta_{\alpha\gamma^{'}}-k_{\alpha}k_{\gamma^{'}})
(\eta_{\alpha^{'}\gamma} \eta_{\alpha^{''} \gamma^{''}}
+\eta_{\alpha^{'}\gamma^{''}} \eta_{\alpha^{''}\gamma }
+\eta_{\alpha^{'}\alpha^{''}} \eta_{\gamma\gamma^{''} })\nn\\
&+&(k^2 \eta_{\alpha\gamma^{''}}-k_{\alpha}k_{\gamma^{''}})
(\eta_{\alpha^{'}\gamma} \eta_{\alpha^{''} \gamma^{'}}
+\eta_{\alpha^{'}\gamma^{'}} \eta_{\alpha^{''}\gamma }+
\eta_{\alpha^{'}\alpha^{''}} \eta_{\gamma \gamma^{'}})~\}\nn\\
             -{ 1\over 8}\{&+&k_{\gamma}k_{\alpha^{'}}
(\eta_{\alpha\gamma^{'}} \eta_{\alpha^{''} \gamma^{''}}
+\eta_{\alpha\gamma^{''}} \eta_{\alpha^{''} \gamma^{'}}
+\eta_{\alpha\alpha^{''}} \eta_{\gamma^{'}\gamma^{''} })\nn\\
&+&k_{\gamma}k_{\alpha^{''}}
(\eta_{\alpha\gamma^{'}} \eta_{\alpha^{'} \gamma^{''}}
+\eta_{\alpha\gamma^{''}} \eta_{\alpha^{'}\gamma^{'} }
+\eta_{\alpha\alpha^{'}} \eta_{\gamma^{'}\gamma^{''} })\nn\\
&+&k_{\gamma}k_{\gamma^{'}}
(\eta_{\alpha\alpha^{'}} \eta_{\alpha^{''} \gamma^{''}}
+\eta_{\alpha\alpha^{''}} \eta_{\alpha^{'} \gamma^{''}}
+\eta_{\alpha\gamma^{''}} \eta_{\alpha^{'}\alpha^{''} })\nn\\
&+&k_{\gamma}k_{\gamma^{''}}
(\eta_{\alpha\alpha^{'}} \eta_{\alpha^{''} \gamma^{'}}
+\eta_{\alpha\alpha^{''}} \eta_{\alpha^{'}\gamma^{'} }
+\eta_{\alpha\gamma^{'}} \eta_{\alpha^{'}\alpha^{''} })~\}\nn\\
   +{ 1\over 4}\{&+&\eta_{\alpha\gamma} (k_{\alpha^{'}}k_{\gamma^{'}}
\eta_{\alpha^{''} \gamma^{''}} + k_{\alpha^{'}}k_{\gamma^{''}}
\eta_{\alpha^{''} \gamma^{'}} + k_{\alpha^{''}}k_{\gamma^{'}}
\eta_{\alpha^{'} \gamma^{''}} \nn\\
&+&k_{\alpha^{''}}k_{\gamma^{''}}
\eta_{\alpha^{'} \gamma^{'}} + k_{\alpha^{'}}k_{\alpha^{''}}
\eta_{ \gamma^{'}\gamma^{''}} +k_{\gamma^{'}}k_{\gamma^{''}}
\eta_{\alpha^{'}\alpha^{''} })~\}.
\eeqa
This expression can be used to calculate divergences. Indeed,
\beqa
k_{\alpha^{'}}H^{~'}_{\alpha\alpha^{'}\alpha^{''}\gamma\gamma^{'}\gamma^{''}}(k)=
{1\over 8}\{&+&(k^2 \eta_{\alpha\alpha^{''}} -k_{\alpha}k_{\alpha^{''}})
(k_{\gamma} \eta_{\gamma^{'} \gamma^{''}}
+k_{\gamma^{'}} \eta_{\gamma \gamma^{''}}
+k_{\gamma^{''}} \eta_{\gamma\gamma^{'} })\nn\\
&+&(k^2 \eta_{\alpha\gamma^{'}}-k_{\alpha}k_{\gamma^{'}})
(k_{\gamma} \eta_{\alpha^{''} \gamma^{''}}
+k_{\gamma^{''}} \eta_{\alpha^{''}\gamma }
+k_{\alpha^{''}} \eta_{\gamma\gamma^{''} })\nn\\
&+&(k^2 \eta_{\alpha\gamma^{''}}-k_{\alpha}k_{\gamma^{''}})
(k_{\gamma} \eta_{\alpha^{''} \gamma^{'}}
+k_{\gamma^{'}} \eta_{\alpha^{''}\gamma }+
k_{\alpha^{''}} \eta_{\gamma \gamma^{'}})~\}\nn\\
             -{ 1\over 8}\{&+&k^{2}k_{\gamma}
(\eta_{\alpha\gamma^{'}} \eta_{\alpha^{''} \gamma^{''}}
+\eta_{\alpha\gamma^{''}} \eta_{\alpha^{''} \gamma^{'}}
+\eta_{\alpha\alpha^{''}} \eta_{\gamma^{'}\gamma^{''} })\nn\\
&+&k_{\gamma}k_{\alpha^{''}}
(2k_{\gamma^{''}}\eta_{\alpha\gamma^{'}}
+2k_{\gamma^{'} }\eta_{\alpha\gamma^{''}}
+k_{\alpha} \eta_{\gamma^{'}\gamma^{''} })\nn\\
&+&k_{\gamma}k_{\gamma^{'}}
(k_{\alpha} \eta_{\alpha^{''} \gamma^{''}}
+2k_{\gamma^{''}}\eta_{\alpha\alpha^{''}} )
+k_{\gamma}k_{\gamma^{''}}k_{\alpha} \eta_{\alpha^{''} \gamma^{'}} ~\}\nn\\
  + { 1\over 4}\{&+&\eta_{\alpha\gamma} (k^{2}k_{\gamma^{'}}
\eta_{\alpha^{''} \gamma^{''}} + k^{2}k_{\gamma^{''}}
\eta_{\alpha^{''} \gamma^{'}} +k^{2} k_{\alpha^{''}}
\eta_{ \gamma^{'}\gamma^{''}} +3k_{\alpha^{''}}k_{\gamma^{'}}k_{\gamma^{''}} ~\}\nn
\eeqa
or using the operator
$H_{\alpha\gamma}= k^2 \eta_{\alpha\gamma} - k_{\alpha}k_{\gamma}$ one can get
\beqa
k_{\alpha^{'}}H^{~'}_{\alpha\alpha^{'}\alpha^{''}\gamma\gamma^{'}\gamma^{''}}(k)=
{1\over 8}\{&+&H_{\alpha\alpha^{''}}~
(k_{\gamma} \eta_{\gamma^{'} \gamma^{''}}
+k_{\gamma^{'}} \eta_{\gamma \gamma^{''}}
+k_{\gamma^{''}} \eta_{\gamma\gamma^{'} })\nn\\
&+&H_{\alpha\gamma^{'}} ~
(k_{\gamma} \eta_{\alpha^{''} \gamma^{''}}
+k_{\gamma^{''}} \eta_{\alpha^{''}\gamma }
+k_{\alpha^{''}} \eta_{\gamma\gamma^{''} })\nn\\
&+&H_{\alpha\gamma^{''}} ~
(k_{\gamma} \eta_{\alpha^{''} \gamma^{'}}
+k_{\gamma^{'}} \eta_{\alpha^{''}\gamma }+
k_{\alpha^{''}} \eta_{\gamma \gamma^{'}})~\}\nn\\
             -{ 1\over 8}\{&+&H_{\alpha\alpha^{''}} ~
              k_{\gamma} \eta_{\gamma^{'}\gamma^{''} }+ H_{\alpha\gamma^{'}} ~
              k_{\gamma} \eta_{\alpha^{''} \gamma^{''}}+H_{\alpha\gamma^{''}}~
              k_{\gamma} \eta_{\alpha^{''} \gamma^{'}}\nn\\
&+&k_{\gamma}k_{\alpha^{''}}
(2k_{\gamma^{''}}\eta_{\alpha\gamma^{'}}
+2k_{\gamma^{'} }\eta_{\alpha\gamma^{''}}
+2k_{\alpha} \eta_{\gamma^{'}\gamma^{''} })\nn\\
&+&k_{\gamma}k_{\gamma^{'}}
(2k_{\alpha} \eta_{\alpha^{''} \gamma^{''}}
+2k_{\gamma^{''}}\eta_{\alpha\alpha^{''}} )
+2k_{\gamma}k_{\gamma^{''}}k_{\alpha} \eta_{\alpha^{''} \gamma^{'}} ~\}\nn\\
  + { 1\over 4}\{&+&\eta_{\alpha\gamma} (k^{2}k_{\gamma^{'}}
\eta_{\alpha^{''} \gamma^{''}} + k^{2}k_{\gamma^{''}}
\eta_{\alpha^{''} \gamma^{'}} +k^{2} k_{\alpha^{''}}
\eta_{ \gamma^{'}\gamma^{''}} +3k_{\alpha^{''}}k_{\gamma^{'}}k_{\gamma^{''}} ~\}\nn
\eeqa
and canceling the identical terms we shall get
\beqa
k_{\alpha^{'}}H^{~'}_{\alpha\alpha^{'}\alpha^{''}\gamma\gamma^{'}\gamma^{''}}(k)=
{1\over 8}\{&+&H_{\alpha\alpha^{''}}
(k_{\gamma^{'}} \eta_{\gamma \gamma^{''}}
+k_{\gamma^{''}} \eta_{\gamma\gamma^{'} })\nn\\
&+&H_{\alpha\gamma^{'}}
(k_{\gamma^{''}} \eta_{\alpha^{''}\gamma }
+k_{\alpha^{''}} \eta_{\gamma\gamma^{''} })\nn\\
&+&H_{\alpha\gamma^{''}}
(k_{\gamma^{'}} \eta_{\alpha^{''}\gamma }+
k_{\alpha^{''}} \eta_{\gamma \gamma^{'}})~\}\nn\\
             -{ 1\over 4}\{
&+&k_{\gamma}k_{\alpha^{''}}
(k_{\gamma^{''}}\eta_{\alpha\gamma^{'}}
+k_{\gamma^{'} }\eta_{\alpha\gamma^{''}})
+k_{\gamma}k_{\gamma^{'}}
k_{\gamma^{''}}\eta_{\alpha\alpha^{''}} ~\}\nn\\
   +{ 1\over 4}\{&+&H_{\alpha\gamma}
   k_{\gamma^{'}}\eta_{\alpha^{''} \gamma^{''}}+H_{\alpha\gamma}
   k_{\gamma^{''}}\eta_{\alpha^{''} \gamma^{'}}+H_{\alpha\gamma}  k_{\alpha^{''}}
\eta_{ \gamma^{'}\gamma^{''}}\nn\\
&+&3\eta_{\alpha\gamma}k_{\alpha^{''}}k_{\gamma^{'}}k_{\gamma^{''}} ~\}.\nn
\eeqa
Again collecting terms we shall get the final expression:
\beqa
k_{\alpha^{'}}H^{~'}_{\alpha\alpha^{'}\alpha^{''}\gamma\gamma^{'}\gamma^{''}}(k)=
{1\over 8}\{&+&H_{\alpha\alpha^{''}}
(k_{\gamma^{'}} \eta_{\gamma \gamma^{''}}
+k_{\gamma^{''}} \eta_{\gamma\gamma^{'} })\nn\\
&+&H_{\alpha\gamma^{'}}
(k_{\gamma^{''}} \eta_{\alpha^{''}\gamma }
+k_{\alpha^{''}} \eta_{\gamma\gamma^{''} })\nn\\
&+&H_{\alpha\gamma^{''}}
(k_{\gamma^{'}} \eta_{\alpha^{''}\gamma }+
k_{\alpha^{''}} \eta_{\gamma \gamma^{'}})~\} \\
             -{ 1\over 4}\{
&+&k_{\gamma}k_{\alpha^{''}}
(k_{\gamma^{''}}\eta_{\alpha\gamma^{'}}
+k_{\gamma^{'} }\eta_{\alpha\gamma^{''}})
+k_{\gamma}k_{\gamma^{'}}
k_{\gamma^{''}}\eta_{\alpha\alpha^{''}}
-3\eta_{\alpha\gamma}k_{\alpha^{''}}k_{\gamma^{'}}k_{\gamma^{''}} ~\}\nn\\
   +{ 1\over 4}\{&+&H_{\alpha\gamma}
   (k_{\gamma^{'}}\eta_{\alpha^{''} \gamma^{''}}+
   k_{\gamma^{''}}\eta_{\alpha^{''} \gamma^{'}}+ k_{\alpha^{''}}
\eta_{ \gamma^{'}\gamma^{''}})~\}\nn,
\eeqa
which has been used in the main text.

\vfill

\begin{thebibliography}{99}

\bibitem{yang} C.N.Yang and R.L.Mills. "Conservation of Isotopic
Spin and Isotopic Gauge Invariance". Phys.\ Rev.\ {\bf 96} (1954) 191

\bibitem{chern}S.S.Chern. {\it Topics in Defferential Geometry},
Ch. III "Theory of Connections"\\
(The Institute for Advanced Study, Princeton, 1951)


\bibitem{Savvidy:2005fi}
G.~Savvidy,
\emph{Non-Abelian tensor gauge fields: Generalization of Yang-Mills theory,}
Phys.\ Lett.\ B {\bf 625} (2005) 341
[arXiv:hep-th/0509049]


\bibitem{Savvidy:2005zm}
  G.~Savvidy,
 \emph{Non-abelian tensor gauge fields. I,}
  Int.\ J.\ Mod.\ Phys.\ A {\bf 21} (2006) 4931;
G.~Savvidy,
\emph{Generalization of Yang-Mills theory: Non-Abelian tensor gauge fields and
higher-spin extension of standard model},
arXiv:hep-th/0505033.

\bibitem{Savvidy:2005ki}
  G.~Savvidy,
  \emph{Non-abelian tensor gauge fields. II,}
  Int.\ J.\ Mod.\ Phys.\ A {\bf 21} (2006) 4959;
G.~Savvidy,
\emph{Non-Abelian tensor gauge fields and extended current algebra: Generalization
of Yang-Mills theory,}
arXiv:hep-th/0510258.


\bibitem{fierz}M.~Fierz. \emph{\"Uber die relativistische Theorie
kr\"aftefreier Teilchen mit beliebigem Spin},  Helv.\ Phys.\ Acta.\  {\bf 12} (1939) 3.

\bibitem{fierzpauli}
M.~Fierz and W.~Pauli.  \emph{On Relativistic Wave Equations for
Particles of Arbitrary Spin in an Electromagnetic Field}, Proc.\ Roy.\ Soc.\  {\bf A173} (1939) 211.




\bibitem{yukawa1}H.Yukawa, \emph{ Quantum Theory of Non-Local Fields.
Part I. Free Fields}, Phys.\ Rev. {\bf 77} (1950) 219 ;~
M.~Fierz, \emph{ Non-Local Fields}, Phys.\ Rev. {\bf 78} (1950) 184



\bibitem{wigner}E.~Wigner, \emph{Invariant Quantum Mechanical Equations of Motion}, in
{\it Theoretical Physics ed. A.Salam}
(International Atomic Energy, Vienna, 1963) p 59


\bibitem{schwinger}J.Schwinger,
\emph{ Particles, Sourses, and Fields}
(Addison-Wesley, Reading, MA, 1970)

\bibitem{Weinberg:1964cn}
S.~Weinberg,
\emph{Feynman Rules For Any Spin},
Phys.\ Rev.\  {\bf 133} (1964) B1318.

\bibitem{chang}S.~J.~Chang, \emph{Lagrange Formulation for Systems with Higher Spin},
Phys.Rev. {\bf 161} (1967) 1308

\bibitem{singh}L.~P.~S.~Singh and C.~R.~Hagen, \emph{ Lagrangian formulation for
arbitrary spin. I. The boson case},
Phys.\ Rev.\ {\bf D9} (1974) 898

\bibitem{singh1}L.~P.~S.~Singh and C.~R.~Hagen, \emph{Lagrangian formulation for
arbitrary spin. II. The fermion case},
Phys.\ Rev.\ {\bf D9} (1974) 898, 910

\bibitem{fronsdal}
C.Fronsdal,
\emph{Massless fields with integer spin},
Phys.Rev. {\bf D18} (1978) 3624


\bibitem{berends}
F.~A.~Berends, G.~J.~H~Burgers and H.~Van Dam,
``On the Theoretical problems in Constructing Interactions Involving
Higher-Spin Massless Particles,''
Nucl.\ Phys.\ B {\bf 260} (1985) 295.


\bibitem{Bengtsson:1983pd}
A.~K.~Bengtsson, I.~Bengtsson and L.~Brink,
``Cubic Interaction Terms For Arbitrary Spin,''
Nucl.\ Phys.\ B {\bf 227} (1983) 31;
A.~K.~Bengtsson, I.~Bengtsson and L.~Brink,
``Cubic Interaction Terms For Arbitrarily Extended Supermultiplets,''
Nucl.\ Phys.\ B {\bf 227} (1983) 41.

\bibitem{Savvidy:2005vm}
G.~Savvidy,
\emph{Non-Abelian tensor gauge fields and higher-spin extension of standard
model,} Fortschr.\ Phys.\ {\bf 54} (2006) 472
[arXiv:hep-th/0512012];
G.~Savvidy and T.~Tsukioka,
\emph{Gauge invariant Lagrangian for non-Abelian tensor gauge fields of fourth
rank,} Prog.\ Theor.\ Phys.\  {\bf 117} (2007) 729
  [arXiv:hep-th/0512344].

\bibitem{Savvidy:2006qf}
  G.~Savvidy,
\emph{Non-Abelian tensor gauge fields: Enhanced symmetries,}
  arXiv:hep-th/0604118.

\bibitem{Guttenberg:2008qe}
  S.~Guttenberg and G.~Savvidy,
 \emph{Schwinger-Fronsdal Theory of Abelian Tensor Gauge Fields,}
  SIGMA 4 (2008)  061 [arXiv:0804.0522].


\bibitem{Barrett:2007nn}
  J.~K.~Barrett and G.~Savvidy,
\emph{A dual lagrangian for non-Abelian tensor gauge fields,}
  Phys.\ Lett.\  B {\bf 652} (2007) 141
   [arXiv:0704.3164];
  S.~Guttenberg and G.~Savvidy,
\emph{Duality transformation of non-Abelian tensor gauge fields,}
  Mod.\ Phys.\ Lett.\  A {\bf 23} (2008) 999
  [arXiv:0801.2459].

\bibitem{Konitopoulos:2007hw}
  S.~Konitopoulos and G.~Savvidy,
 \emph{Propagating modes of non-Abelian tensor gauge field of second rank,}
  J.\ Phys.\ A  {\bf 41} (2008) 355402
  [arXiv:0706.0762].







\bibitem{Savvidy:2008ks}
  G.~Savvidy,
  \emph{Connection between non-Abelian tensor gauge fields and open strings,}
  J.\ Phys.\ A  {\bf 42} (2009) 065403;
  I.~Antoniadis and G.~Savvidy,
  \emph{Scattering of charged tensor bosons in gauge and superstring theories,}
  arXiv:0907.3553.

\bibitem{Coleman:1967ad}
  S.~R.~Coleman and J.~Mandula,
 \emph{All Possible Symmetries Of The S Matrix,}
  Phys.\ Rev.\  {\bf 159} (1967) 1251.

\bibitem{Gross:1988ue}
  D.~J.~Gross,
 \emph{High-Energy Symmetries Of String Theory,}
  Phys.\ Rev.\ Lett.\  {\bf 60} (1988) 1229.

\bibitem{Haag:1974qh}
  R.~Haag, J.~T.~Lopuszanski and M.~Sohnius,
 \emph{All Possible Generators Of Supersymmetries Of The S Matrix,}
  Nucl.\ Phys.\  B {\bf 88} (1975) 257.

\bibitem{Wess:1992cp}
  J.~Wess and J.~Bagger,
 \emph{Supersymmetry and supergravity,}
{\it  Princeton, USA: Univ. Pr. (1992) 259 p}






\end{thebibliography}
\end{document}